\documentclass[11pt,a4paper]{article}
\pdfoutput=1 

\usepackage[utf8x]{inputenc}
\usepackage{graphicx}

\usepackage[T1]{fontenc}
\usepackage{amssymb,amsmath,amsfonts}
\usepackage{stackrel}
\usepackage{float}
\usepackage{dsfont}
\usepackage{jheppub}

\newcommand{\ms}[1]{\textrm{\tiny $\rm #1$}}

\newcommand{\be}{\begin{equation}}
\newcommand{\ee}{\end{equation}}
\newcommand{\bse}{\begin{subequations}}
\newcommand{\ese}{\end{subequations}}
\newcommand{\ba}{\begin{eqnarray}}
\newcommand{\ea}{\end{eqnarray}}
\renewcommand{\(}{\left(}
\renewcommand{\)}{\right)}

\newcommand{\6}{\partial}

\newcommand{\bea}{\begin{eqnarray}}
\newcommand{\eea}{\end{eqnarray}}

\newcommand{\ab}{\alpha\beta}
\newcommand{\mn}{\mu\nu}

\newcommand{\gb}{\bar\gamma}

\newcommand{\wT}{\hat{\mathcal{T}}}
\newcommand{\wH}{\hat{\mathcal{H}}}
\newcommand{\wA}{\hat{\mathcal{A}}}
\newcommand{\wE}{\hat{\mathcal{E}}}
\newcommand{\wP}{\hat{\mathcal{P}}}

\begin{document}

\title{Semi-Holography for Heavy Ion Collisions: Self-Consistency and First Numerical Tests}

\author[]{ Ayan Mukhopadhyay, Florian Preis, Anton Rebhan and Stefan A.~Stricker}

\affiliation[]{Institut f\"{u}r Theoretische Physik, Technische Universit\"{a}t Wien,\\
Wiedner Hauptstr.~8-10, A-1040 Vienna, Austria}

\emailAdd{ayan@hep.itp.tuwien.ac.at}
\emailAdd{rebhana@hep.itp.tuwien.ac.at}
\emailAdd{fpreis@hep.itp.tuwien.ac.at}
\emailAdd{stricker@hep.itp.tuwien.ac.at}

\abstract{
We present an extended version of a recently proposed semi-holographic model for heavy-ion collisions, 
which includes self-consistent
couplings between the Yang-Mills fields of the Color Glass Condensate framework and an infrared AdS/CFT sector, such as to guarantee the existence of a conserved energy-momentum tensor for the combined system that is local in space and time, which we also construct explicitly. Moreover, we include
a coupling of the topological charge density in the glasma to the same of the holographic infrared CFT. 
The semi-holographic approach makes it possible to combine CGC initial conditions and weak-coupling glasma field equations
with a simultaneous evolution of a strongly coupled infrared sector describing the soft gluons radiated by hard partons.
As a first numerical test of the semi-holographic model we study the dynamics
of fluctuating homogeneous color-spin-locked Yang-Mills fields when coupled to
a homogeneous and isotropic energy-momentum tensor of the holographic IR-CFT,
and we find rapid convergence of the iterative numerical procedure suggested earlier.
}

\maketitle


\section{Introduction}

Describing the full time evolution and equilibration process of the fireball created in ultrarelativistic heavy-ion collisions is an extremely difficult task due to the interplay of perturbative and non-perturbative phenomena.
Tracing the full evolution appears to require a patchwork of different effective  theories, each  designed to describe a certain stage of the evolution and only applicable  to specific physical observables.

According to the color glass condensate (CGC) framework \cite{Gelis:2010nm},
at very early times  $\tau\sim 1/Q_s$ 
the system is dominated by gluons
with typical momenta of the order of the semi-hard saturation scale $Q_s$, which are
relatively weakly coupled (when $Q_s$ is large), $\alpha_s(Q_s)\ll1$, but which have high occupation numbers $\sim 1/\alpha_s$.
This admits a
description in terms of semiclassical Yang-Mills fields,
also called the glasma \cite{Lappi:2006fp}, which
has been studied by numerical solutions of classical Yang-Mills equations \cite{Romatschke:2006nk,Gelis:2013rba,Berges:2013fga
}.
As the system undergoes rapid longitudinal expansion and the occupation number of the semi-hard gluons drops, 
soft gluons are emitted abundantly, whose dynamics is characterised by a significantly larger coupling $\alpha_s(\mu< Q_s)$.
This growing soft sector is presumed to play a crucial role in the evolution of the whole system.

While the rapid thermalization suggested by experimental data might be captured by an effective
kinetic theory description extrapolated to strong coupling \cite{Kurkela:2015qoa},
insight into isotropization and thermalization of a strongly coupled quantum field theory
can be comparatively directly gained by gauge/gravity duality.
The latter maps the thermalization process to
black hole formation in asymptotically anti-de Sitter space
\cite{DeWolfe:2013cua,Chesler:2013lia}.  
By comparing results from hydrodynamic simulations with the full holographic evolution it was found that the system is accurately described by hydrodynamics although large viscous corrections are still present. Later the system isotropizes  and reaches local thermal equilibrium \cite{Chesler:2010bi, Heller:2011ju}. (For an extensive list of references see the recent review \cite{Chesler:2015lsa}.)

The mechanism how the initial state makes a transition to a state that is amenable to a hydrodynamic description is however different for weak and strong coupling (although quantitatively there might
exist a smooth interpolation \cite{Keegan:2015avk}).
At asymptotically high energies and parametrically small coupling the thermalization pattern is of the bottom-up type \cite{Baier:2000sb,Kurkela:2011ub},
where first the soft gluons form a thermal bath which then draws energy from the hard modes.
On the other hand,
at infinite  coupling the thermalization process as obtained via the gauge/gravity correspondence is top-down, meaning that highly  energetic modes reach their thermal distribution first \cite{Balasubramanian:2011ur}. 
By considering corrections to
the infinite-coupling limit it was shown in \cite{Steineder:2012si, Stricker:2013lma} that there indeed
exists a transition between the two behaviors at intermediate coupling.  

A description of the whole evolution from within a
single framework and from first principles 
is elusive at this point. So far most studies have either utilized  solely a weakly coupled approach  or a strongly coupled one, but in order to understand the evolution of the colliding matter better, one needs to combine the  different effective descriptions for the different stages  as well as weak and strong coupling phenomena.
In recent years various attempts in this direction have been  made. 
For example,   in \cite{vanderSchee:2013pia}  different effective descriptions were patched together. The far-from-equilibrium initial  stage was simulated by colliding shock waves using  numerical AdS/CFT, whose results were used as an input for the hydrodynamic evolution, which subsequently  served as an input for the kinetic theory describing the low density hadronic stage. 

However, the initial conditions set up for AdS/CFT calculations, even in state-of-the-art shock-wave collisions, 
have no clear connection to those
derived directly from (weak-coupling) QCD in the (nonperturbative) CGC framework. It would seem highly desirable
to be able to connect the two approaches and follow the combined evolution of weakly coupled semi-hard gluons
and a strongly interacting soft sector.

The first semi-holographic  proposals  for  describing  interactions between a weakly coupled  and a strongly coupled sector, where the strongly coupled  sector is  described by a holographic dual, were made in the context of non-Fermi liquids
\cite{Faulkner:2010tq, Mukhopadhyay:2013dqa}.
In the context of heavy-ion collisions one attempt to combine weak and strong coupling was made in a so-called hybrid approach to describe the energy loss of jets moving through a  strongly coupled medium \cite{Casalderrey-Solana:2014bpa}, where the soft in-medium effects were modeled by using insights from gauge/gravity duality. However, in this case no back reaction of the soft medium to the hard partons was taken into account. 

A different route was recently taken in \cite{Iancu:2014ava} where a semi-holographic model for thermalization in heavy-ion collisions was presented. 
This model is able to incorporate the interaction between weakly coupled hard and strongly coupled soft modes but so far lacks verification by carrying out a concrete calculation. 
The goal of this work is to refine and test the  proposal of \cite{Iancu:2014ava}
with regard to its internal self-consistency and numerical practicability
(while not yet attempting to shed new light on
thermalization in heavy-ion collisions).

The outline of this paper is as follows. In Sec.~\ref{sec:semiholo} we review and also extend the semi-holographic model of \cite{Iancu:2014ava}. 
Specifically, we present a reformulation of the model which permits the definition of a locally conserved energy-momentum tensor of the combined system of hard and soft degrees of freedom.
In Sec.~\ref{sec:toymodel} we present a simple test case
for semi-holographic glasma evolution. We truncate the couplings of the model developed in Sec.~\ref{sec:semiholo}, such that only gravitational degrees of freedom are excited in the holographic sector and moreover impose homogeneity and isotropy in both the hard and the soft sector. As a consequence the (time-dependent) gravity dual of this test case can be treated
analytically, but due to the Birkhoff theorem does not allow for
dynamical black-hole formation or perturbation. Thus, the soft degrees of freedom captured by the dual strongly coupled field theory cannot equilibrate with the hard degrees of freedom, i.e., this simple test case is too restricted for permitting to study thermalization.
It however allows us to
carry out a first proof-of-principle calculation, where we can verify the convergence of an iterative numerical solution of
the semi-holographic set-up. Section~\ref{sec:conclusions} contains our conclusions and an outlook.

\section{Action and energy-momentum tensor of the semi-holographic set-up}\label{sec:semiholo}

Following \cite{Iancu:2014ava}, in this section we construct an action 
where the description of a semi-hard gluon (UV) sector and that of a strongly-interacting soft gluon (IR) 
sector can be coupled self-consistently. In this paper, we will treat
the semi-holographic model phenomenologically including minimalistic hard-soft couplings
only, without attempting a derivation from first principles; a rigorous framework
for a possible derivation has been recently proposed in \cite{Behr:2015aat}.

The assumptions that provide the basis of the semi-holographic model for heavy-ion collisions in \cite{Iancu:2014ava} are as follows:
\begin{enumerate}
\item the over-occupied semi-hard (UV) gluons allow for a description in terms of a classical Yang--Mills theory as in the glasma model \cite{Gelis:2010nm,Lappi:2006fp},
\item the infrared sector is described by a conformal field theory (IR-CFT) which is approximated by the strong-coupling large-$N$ limit 
such that a classical dual holographic description is possible,
\item the IR-CFT is marginally deformed such that its marginal couplings become functionals of the color fields of the glasma, while the marginal operators of the infrared conformal field theory (IR-CFT) appear as self-consistent fields which modify the classical Yang-Mills dynamics of the color fields of the glasma, and
\item the hard-soft couplings describing the mutual feedback between the IR-CFT and the color fields of the glasma involve local gauge-invariant operators of both sectors, and the coupling constants are given by
a dimensionless number times $Q_s^{-4}$, with $Q_s$ being the saturation scale of the colliding nuclei.
\end{enumerate}
 
In addition to the proposal in \cite{Iancu:2014ava} we demand a well defined action principle that allows for the construction of a conserved energy-momentum tensor for the combined system living in flat space.

For the action describing the interactions between the hard and soft modes, which satisfies all the criteria mentioned above, we propose
\begin{equation}
S=S_{\mathrm{YM}}+W_{\mathrm{CFT}}[g^{({\rm b})}_{\mn},\phi^{({\rm b})},\chi^{({\rm b})}]\,,\label{eq:action}
\end{equation}
which we will discuss now in detail. 

We start with the Yang-Mills part of the action and for the time being, consider a classical Yang-Mills theory living on an arbitrary non-dynamical background metric $g^\mathrm{YM}_{\mu\nu}$, which in the end will be set to the flat Minkowski metric. This is for technical purposes only because it is convenient for identifying the tensorial properties of the variables to be defined below.
The Yang-Mills action $S_{\mathrm{YM}}$ in this background is given by 
\be
S_{\mathrm{YM}}=-\int \mathrm{d}^4 x\,\sqrt{-g^{\mathrm{YM}}}\;h,\;\qquad h:=\frac{1}{4N_c}\mathrm{Tr} \(F_{\mn}F^{\mn}\).
\ee
We use the normalisation  $\mathrm{Tr}(T^a T^b)=N_c \delta^{ab}$ for the generators of the $SU(N_c)$ gauge group together with the  standard convention for the covariant derivative $D_{\mu}=\nabla_\mu-ig A_\mu^a T^a$, where $\nabla_\mu$ is the Levi-Civita connection with respect to $g^{\mathrm{YM}}_{\mn}$.

The gauge invariant operators of the UV sector used to deform the IR-CFT are $h$, the energy-momentum tensor of the Yang-Mills fields, $t_{\mn}$, which takes the standard form
 \be\label{UVemt}
 t_{\mn}=\frac{1}{N_c}\mathrm{Tr}\(F_{\mu\alpha}F_{\nu}^{~\alpha}-\frac{1}{4}g^{\mathrm{YM}}_{\mn}F_{\ab}F^{\ab}\)\,,
 \ee
and the Yang-Mills Pontryagin density
\be
a:=\frac{1}{4\sqrt{-g^\mathrm{YM}}N_c}\mathrm{Tr} \(F_{\mn} \tilde{F}^{\mn}\).
\ee
Notice that $a$, $h$ as well as $t_{\mn}$ are functionals of the background metric $g^\mathrm{YM}_{\mu\nu}$.

The second part of the action \eqref{eq:action}, $W_{\mathrm{CFT}}$, is the generating functional of the IR-CFT, provided by the on-shell gravitational action
of its gravity dual. The gravitational dynamics will be described by Einstein's gravity with asymptotically AdS boundary conditions minimally coupled to matter.\footnote{It is to be noted that the AdS radius $l$ does not explicitly appear in the IR-CFT variables in the strong coupling and  large $N$ limit. The $\alpha'/l^2$ parameter which gives higher derivative corrections to Einstein's gravity is identified with $1/\sqrt{\lambda}$ (up to a numerical factor) where $\lambda$ is the analogue of the 't Hooft coupling of the IR-CFT. This disappears when we take the limit $\lambda \rightarrow \infty$. The factor $(4\pi G_5)/l^3$ is proportional to $\hat{N}_c^2$, where $\hat{N}_c$ is the analogue of the rank of the gauge group of the IR-CFT. This should be parametrically of the same magnitude as the $N_c$ of QCD---the relative numerical factor can be absorbed in the definitions of the hard-soft coupling constants given below.} 
However, at this point we can be completely agnostic about the IR-CFT generating functional $W_{\rm CFT}$, 
meaning that we do not need to know its explicit form in terms of the elementary fields. We only need that $W_{\rm CFT}$ describes a marginally deformed CFT in presence of three background sources, namely $g^{\rm (b)}_{\mn}$, $\phi^{\rm (b)}$ and $\chi^{\rm (b)}$, which couple to the three marginal operators of dimension $4$, namely $\mathcal{T}_{\mn}$ (the energy-momentum tensor of the CFT), $\mathcal{H}$ (the glueball/Lagrangian density operator of the CFT) and $\mathcal{A}$ (the topological charge density operator of the CFT), respectively\footnote{To be consistent with the notation in \cite{Iancu:2014ava}, we denote IR-CFT operators by calligraphic capital letters.}. The bulk fields dual to these operators are the metric $G_{MN}$, the dilaton $\phi$, and the axion $\chi$. For example, the IR-CFT energy-momentum tensor can be obtained from the asymptotic expansion of the bulk metric $G_{MN}$ in Fefferman-Graham gauge, which takes the form
\bea\label{asymp}
G_{\rho\rho}&=&\frac{l^2}{\rho^2},\quad\quad G_{\rho\mu}=0,\nonumber\\
G_{\mn}&=&\frac{l^2}{\rho^2}\(g^{({\rm b})}_{\mn}+\cdot\cdot\cdot +\rho^4\(\frac{4\pi G_5}{l^3} \mathcal{T}_{\mn}+ X_{\mn}\)+\mathcal{O}\left(\rho^6\right)\),
\eea
where $\rho$ denotes the holographic radial coordinate, with $\rho=0$ being the location of the conformal boundary 
of the bulk spacetime. The tensor $X_{\mn}$ above is also a local functional of the boundary metric $g^{({\rm b})}_{\mn}$, is non-trivial when the latter is curved, and is explicitly known (see \cite{deHaro:2000xn}). Likewise the boundary values of the axion and dilaton are $\phi^{\rm (b)}$ and $\chi^{\rm (b)}$, respectively.

We also have to specify how $g^{\rm (b)}_{\mn}$, $\phi^{\rm (b)}$ and $\chi^{\rm (b)}$ are determined by the Yang-Mills color fields as appropriate gauge-invariant tensors. At leading order in $Q_s^{
-4}$, the most general forms of these sources are:\footnote{We also require that the full action is CP-invariant so that we can rule out $a$-$\mathcal{H}$ or $h$-$\mathcal{A}$ couplings.}
\begin{subequations}\label{boundarycond}
\bea
g^{({\rm b})}_{\mu\nu}&=&g^\mathrm{YM}_{\mu\nu}+\frac{\gamma}{Q_s^4}t_{\mu\nu},\label{gb}\\
\phi^{({\rm b})}&=& \frac{\beta}{Q_s^ 4}h\label{phib}\;,\qquad \chi^{({\rm b})}= \frac{\alpha}{Q_s^ 4}a,\label{chib}
\eea
\end{subequations}
where $\gamma$, $\beta$ and $\alpha$ are  dimensionless free parameters. The IR-CFT $\mathcal{T}_{\mu\nu}$ is thus determined by the color fields, after assuming appropriate initial conditions and requiring absence of naked singularities in the bulk spacetime. Similarly, all other IR-CFT operators are also determined by the color fields of the glasma via holographic dynamics.

Heuristically, the above form of the bulk fields at the boundary can be understood in terms of renormalisation group (RG) flow, where the scale of the flow is related to the radial direction of the dual gravity theory.
Going along the radial coordinate both the sources (including the background metric) and the single trace operators of the CFT evolve in the dual geometries.
It was shown in \cite{Behr:2015yna,Behr:2015aat} that it can be recast in a precise form of RG flow in which the Ward identities of single trace operators giving local conservation laws preserve their form along the scale evolution, so that the sources become functionals of the ultraviolet degrees of freedom.
 Although this RG flow occurs in a fixed background metric,
the sources emerge as scale-dependent effective dynamical
objects whose evolution is described by a dual diffeomorphism-invariant classical gravity
theory in one higher dimension in the large $N$ limit. 
 In order to include weakly coupled
UV degrees of freedom which cannot be described by a dual classical gravity theory, it is
natural to stop the geometric description of dynamics at a certain scale, in our case the saturation scale $Q_s$
 and assume that
the effective sources and background metric at that scale are functionals of the operators
of the weakly coupled UV fields instead of the UV modes of the strongly coupled holographic
CFT.
Therefore the boundary metric (\ref{gb}) should be rather viewed as an \textit{effective} metric that 
accounts for the marginal deformation of the IR-CFT due to the coupling between the soft and hard modes at the scale $Q_s$.
 In fact, as we will show  below,  the full  energy-momentum tensor that follows from our model lives  in the background metric $g^{\mathrm{YM}}_{\mn}$, which is set to $\eta_{\mu\nu}$ eventually.

The appropriate initial conditions in gravity, in the context of heavy-ion collisions, is that the bulk geometry is pure $AdS_5$ with vanishing dilaton and axion fields \cite{Iancu:2014ava}, reflecting the fact that the initial dynamics is primarily in the hard sector. Thus imposing the boundary values \eqref{boundarycond} on all the bulk fields and demanding regularity at the future horizon, we obtain a unique gravity  solution  which gives a well-defined $W_{\rm CFT}$ that is equal to the dual gravitational on-shell action. It has been observed in \cite{Iancu:2014ava} that the hard-soft couplings do not modify the glasma 
initial 
conditions \cite{Kovner:1995ja, Kovner:1995ts} and this remains true in the extended construction here as well.\footnote{Technically, one may need to make the hard-soft couplings $\alpha$, $\beta$ and $\gamma$ time-dependent, e.g.\ proportional to $\tanh(Q_s\tau)$, such that they start from zero and become almost a constant at a time scale of order $Q_s^{-1}$, the time required by the uncertainty principle for a gluon of virtuality $Q_s^{-1}$ to emit a soft quanta \cite{Iancu:2014ava}.}

By varying the above action (\ref{eq:action}) with respect to the Yang-Mills fields $A_{\mu}$ according to
 \be\label{variation}
 \frac{\delta S}{\delta A_\mu(x)}=\frac{\delta S_{\mathrm{YM} }}{\delta A_{\mu}(x)}+\int\mathrm{d}^4 y\(\frac{\delta W_{\mathrm{CFT}}}{\delta g^{\rm (b)}_{\ab}(y)}\frac{\delta  g^{\rm (b)}_{\ab}(y) }{\delta A_{\mu}(x)}+\frac{\delta W_{\mathrm{CFT}}}{\delta \phi^{\rm (b)}(y)}\frac{\delta  \phi^{\rm (b)}(y)}{\delta A_{\mu}(x)}+\frac{\delta W_{\mathrm{CFT}}}{\delta \chi^{\rm (b)}(y)}\frac{\delta  \chi^{\rm (b)}(y)}{\delta A_{\mu}(x)}\)
 \ee 
 we obtain the equations of motion for the Yang-Mills color fields modified by the marginal operators of the IR-CFT
\begin{equation}
\mathcal{T}^{\alpha\beta}=\frac{2}{\sqrt{-g^{\rm (b)}}}\frac{\delta W_{\mathrm{CFT}}}{\delta g^{\rm (b)}_{\alpha\beta}},\quad \mathcal{H}=\frac{1}{\sqrt{-g^{\rm (b)}}}\frac{\delta W_{\mathrm{CFT}}}{\delta \phi^{\rm (b)}},\quad \mathcal{A}=\frac{1}{\sqrt{-g^{\rm (b)}}}\frac{\delta W_{\mathrm{CFT}}}{\delta \chi^{\rm (b)}}.
\end{equation}

In order to make clear that all terms in the equations of motion indeed have the same tensor structure,
it is useful to define the following new tensorial objects 
\begin{equation}\label{eq:hatIRobjects}
\wT^{\alpha\beta}=\frac{\sqrt{-g^{(b)}}}{\sqrt{-g^{\mathrm{YM}}}}\; \mathcal{T}^{\alpha\beta}, \quad
\wH=\frac{\sqrt{-g^{(b)}}}{\sqrt{-g^{\mathrm{YM}}}}\;\mathcal{H},\quad \wA=\frac{\sqrt{-g^{(b)}}}{\sqrt{-g^{\mathrm{YM}}}}\;\mathcal{A}\;,
\end{equation}
such that the modified Yang-Mills equations read
\begin{equation}\label{eom}
D_\mu F^{\mu\nu}=\frac{\beta}{Q_s^4}D_\mu \left(\wH F^{\mu\nu}\right)+\frac{\alpha}{Q_s^4}\(\partial_\mu\wA\) \,\tilde{F}^{\mu\nu}+\frac{\gamma}{Q_s^4}D_\mu\left(\wT^{\mu\alpha}F_\alpha^{\ \nu}-\wT^{\nu\alpha}F_\alpha^{\ \mu}-\frac{1}{2}\wT^{\alpha\beta}g^{\rm YM}_{\alpha\beta}F^{\mu\nu}\right).
\end{equation}
It is to be noted that $\wT^{\alpha\beta}$, $\wH$ and $\wA$ are tensors living in the background metric $g^{\mathrm{YM}}_{\mu\nu}$, because $g^{\rm (b)}_{\mn}$ is itself such a tensor and the factor $\sqrt{-g^{\rm (b)}}/\sqrt{-g^{\mathrm{YM}}}$ is an invariant scalar under diffeomorphisms.\footnote{Indeed this factor $\sqrt{-g^{\rm (b)}}/\sqrt{-g^{\mathrm{YM}}}$ has a nice geometric interpretation. One can readily check that if $J^\mu$ is a local conserved current in the background metric $g^{\rm (b)}_{\mn}$, meaning that it satisfies $\nabla_{{\rm (b)}\mu}J^\mu = 0$ with $\nabla_{{\rm (b)}}$ being the covariant derivative constructed from $g^{(b)}$, then $\hat{J}^\mu = \sqrt{-g^{\rm (b)}}/\sqrt{-g^{\mathrm{YM}}} J^\mu$ is a conserved current in the background $g^{\mathrm{YM}}_{\mu\nu}$ satisfying $\nabla_{\mu}\hat{J}^\mu = 0$ with $\nabla_{\mu}$ being the covariant derivative constructed from $g^{\rm YM}_{\mn}$.} 
In the above equations indices are still lowered and raised with the metric 
$g_{\mu\nu}^\mathrm{YM}$ and its inverse respectively, while $\mathcal{T}^{\alpha\beta}$ is related to $\mathcal{T}_{\mn}$ in \eqref{asymp} via $\mathcal{T}^{\alpha\beta} = g^{{\rm (b)}\alpha\mu}\mathcal{T}_{\mn}g^{{\rm (b)}\nu\beta}$.

The full energy-momentum tensor $T^{\mu\nu}$ of the coupled UV-IR model is obtained by varying the action (\ref{eq:action})  with respect to the metric according to
\ba
T^{\mu\nu}&=&\frac{2}{\sqrt{-g^{\mathrm{YM}}}}\Bigg[\frac{\delta S_{\mathrm{YM}}}{\delta  g^{\mathrm{YM}}_{\mn}(x)}\nonumber\\
&&+\int \mathrm{d}^4y\(\frac{\delta W_{\mathrm{CFT}}}{\delta g^{(b)}_{\ab}(y)}\frac{\delta g^{(b)}_{\ab}(y)}{\delta g^{\mathrm{YM}}_{\mn}(x)}+\frac{\delta W_{\mathrm{CFT}}}{\delta \phi^{(b)}(y)}\frac{\delta  \phi^{(b)}(y)}{\delta g^{\mathrm{YM}}_{\mn}(x)}+\frac{\delta W_{\mathrm{CFT}}}{\delta \chi^{(b)}(y)}\frac{\delta  \chi^{(b)}(y)}{\delta g^{\mathrm{YM}}_{\mn}(x)}\)\Bigg] .\quad
\ea
When the Yang-Mills metric is finally set to the Minkowski metric $g^{\mathrm{YM}}_{\mu\nu}=\eta_{\mu\nu}$, this gives
\begin{eqnarray}\label{EMT}
T^{\mu\nu}&=&t^{\mu\nu}\nonumber\\
&\ &+\wT^{\mu\nu}-\frac{\gamma}{Q_s^4N_c}\wT^{\alpha\beta}\left[\mathrm{Tr}(F_\alpha^{\ \mu}F_\beta^{\ \nu})-\frac{1}{2}\eta_{\alpha\beta}\mathrm{Tr}(F^{\mu\rho}F^{\nu}_{\ \rho})+\frac{1}{4}\delta_{(\alpha}^\mu\delta_{\beta)}^\nu\mathrm{Tr}(F^2)\right]\nonumber\\
&\ & -\frac{\beta}{Q_s^4N_c}\wH\,\mathrm{Tr}(F^{\mu\alpha}F^{\nu}_{\ \alpha})-\frac{\alpha}{Q_s^4}\eta^{\mu\nu}\wA\,a.
\end{eqnarray}
In the following we will refer to $t^{\mu\nu}$ as the energy-momentum tensor describing the hard sector, while the remainder of $T^{\mu\nu}$, which is obtained from $W_{\mathrm{CFT}}$, will be ascribed to the soft sector. 
These latter contributions include the effects of interactions between the hard and the soft
sector as apparent from the terms involving products of $\wT^{\alpha\beta}$ and expressions
bilinear in $F_{\mn}$. However, there is no unambiguous separation between
purely soft and soft-hard contributions. Instead of $\wT^{\mu\nu}$ one could
equally well separate off e.g.\ $\mathcal{T}^{\mu\nu}$ or $\mathcal{T}_{\mu\nu}=g^{\rm (b)}_{\mu\alpha}\mathcal{T}^{\alpha\beta}g^{\rm (b)}_{\beta\nu}$ as putative purely soft part,
which would all give slightly different decompositions into a soft energy-momentum tensor
and mixed terms involving also the Yang-Mills field (except when $\gamma=0$,
where the different soft energy-momentum tensors coincide).\footnote{The entropy of the soft sector
can however be tracked by considering the area of the black hole horizon forming in the gravitational dual.}
In Appendix \ref{EMTconserv} we explicitly show that the full energy-momentum tensor is conserved on shell, i.e., if we impose  the equations of motion \eqref{eom} together with the Ward identity {for local conservation of energy and momentum} of the IR-CFT {in the self-consistent background $g^{\rm (b)}_{\mn}$}.\footnote{This Ward identity, which is a generic feature of any field theory that can be coupled to an arbitrary background metric in diffeomorphism invariant manner, also follows from the vector constraints of the dual gravitational theory.}

The existence of a full energy-momentum tensor which is locally conserved in the \textit{actual} background metric for the Yang-Mills dynamics, i.e. $\partial_\mu T^{\mu\nu}=0$, is the primary improvement on the original semi-holographic construction of Ref.~\cite{Iancu:2014ava} that has been achieved here. In order to compare with \cite{Iancu:2014ava}, it is useful 
to  define new variables:
\begin{eqnarray}\label{IRbar}
\overline{\mathcal{T}}^{\alpha\beta}&=&\sqrt{-g^{\mathrm{YM}}}\; \hat{\mathcal{T}}^{\alpha\beta} \,=\, \sqrt{-g^{\mathrm{(b)}}}\; \mathcal{T}^{\alpha\beta} \,=\, 2\frac{\delta W_{\mathrm{CFT}}}{\delta g^{\rm (b)}_{\ab}}, \nonumber\\
\overline{\mathcal{H}}&=&\sqrt{-g^{\mathrm{YM}}}\;\hat{\mathcal{H}}\, =\, \sqrt{-g^{\mathrm{(b)}}}\; \mathcal{H} \,=\, \frac{\delta W_{\mathrm{CFT}}}{\delta \phi^{\rm (b)}},\nonumber\\ \overline{\mathcal{A}}&=&\sqrt{-g^{\mathrm{YM}}}\;\hat{\mathcal{A}} \,=\, \sqrt{-g^{\mathrm{(b)}}}\; \mathcal{A} \, =\,\frac{\delta W_{\mathrm{CFT}}}{\delta \chi^{\rm (b)}},
\end{eqnarray}
which transform as tensorial densities. Using these variables, we can replace the  effective action (\ref{eq:action}) by
\bea\label{effaction}
S_\mathrm{glasma}&=&-\frac{1}{4N_c}\int \mathrm{d}^4x\, \sqrt{-g^{\rm YM}}\,\mathrm{Tr}(F_{\ab} F^{\ab})+\int \mathrm{d}^4 x \,\overline{\mathcal{H}}\,\phi^{\rm (b)}\nonumber\\
&\ &+\int \mathrm{d}^4 x\, \overline{\mathcal{A}} \,\chi^{\rm (b)}
+\frac{1}{2}\int \mathrm{d}^4x \,\overline{\mathcal{T}}^{\mn}g^{\rm (b)}_{\mu\nu},
\eea
which is invariant under diffeomorphisms. Treating $\overline{\mathcal{T}}^{\mn}$, $\overline{\mathcal{A}}$, and $\overline{\mathcal{H}}$ as \textit{independent} IR-CFT variables, we can vary the above action with respect to the Yang-Mills gauge fields to obtain the equations of motion \eqref{eom}, while variation
with respect to $g_{\mn}^{\rm YM}$ yields the full energy-momentum tensor (\ref{EMT}), eventually setting $g_{\mn}^{\rm YM}$ to the flat Minkowski space metric. 

To linear order in the coupling constants (without $\alpha$), 
to which \cite{Iancu:2014ava} had restricted itself,
the modified glasma field equations of \cite{Iancu:2014ava} are reproduced by \eqref{eom}
apart from a factor of $-\frac{1}{2}$ in the term proportional to $\mathcal{T}^{\mn}$ and a factor of $-1$ in the term proportional to $\mathcal{H}$. Beyond linear order in $\gamma$, the crucial difference is the presence of $\sqrt{-g^{\mathrm{(b)}}}$ in
the densities (\ref{IRbar}). While the entire system is residing in flat Minkowski space, the IR-CFT
is effectively living in a nontrivial background $g^{\rm (b)}_{\mn}-\eta_{\mn}\propto\gamma$.
Keeping all orders in $\gamma$ is necessary to have an exactly conserved energy-momentum tensor.\footnote{Although
inconsequential as far as the equations of motion are concerned, for
the possibility to derive the conserved total energy-momentum tensor from \eqref{effaction} it is also
important that the last term in \eqref{effaction} involves the trace of $\overline{\mathcal T}^{\mn}$
with respect to $g_{\mn}^{\rm YM}$ and not only a contraction with $t_{\mn}$.}

As suggested in \cite{Iancu:2014ava}, the equations of motion may  be solved self-consistently by an iterative process. First the classical YM equations with vanishing expectation values for the soft sector operators $\wT^{\alpha\beta},~\wH$ and $\wA$ are solved. 
With this solution  the sources for the gravitational problem are evaluated and its field equations are solved in a second step. With the gravity solution at hand  the IR operators    are extracted and  plugged  back into (\ref{eom}), and the next round of the iteration process can be performed. This is done until convergence for both sectors is reached. At each step in the iteration procedure, the initial conditions in the Yang-Mills and gravity sectors mentioned earlier are held fixed. An important feature of this approach is that the usual ad-hoc initial conditions used for  gravity calculations to model thermalization can now entirely 
be determined by the associated color-glass condensate description as discussed above. 

Since the initial glasma field configurations of a heavy-ion collisions involve strong longitudinal chromo electric and magnetic fields and thus significant Pontryagin density, it will also be of interest to study its coupling to the gravitational axion
of the dual IR-CFT provided by the term involving $\overline{\mathcal{A}}$ in (\ref{effaction}). Indeed, a nontrivial axion field has been introduced previously in static models of strongly coupled anisotropic
super-Yang-Mills plasma \cite{Mateos:2011ix}
with interesting consequences such as a violation of the usual bound on the shear viscosity \cite{Rebhan:2011vd}. The
presence of a nonvanishing Pontryagin density is moreover of central interest to anomalous transport phenomena such as the chiral
magnetic effect \cite{Kharzeev:2015znc}.

\section{A simple test case}\label{sec:toymodel}

In this section, we present the first numerical test of the above semi-holographic setup in a very simple test case of coupling classical Yang-Mills simulations to a holographic description of a strongly coupled soft sector. In this test case the energy-momentum tensor is homogeneous and isotropic, but both the UV and IR degrees of freedom have
nontrivial (0+1-dimensional) dynamics. This allows us
to check for convergence of the proposed iterative scheme for solving both the UV and IR dynamics. Furthermore, we also check that, when the iterative procedure converges, the full energy-momentum tensor (\ref{EMT}) of the combined system is indeed conserved. 

For simplicity, we also switch off the hard-soft couplings $\alpha$ and $\beta$ so that
the dilaton and the axion are not excited, while retaining $\gamma$. The full solution in our test case shows periodic transfers of energy from hard to soft sector and vice versa
without any thermalization due to the fact that the symmetries in the gravitational sector of the test case exclude any propagating degrees of freedom and thus do not allow for dynamical black hole formation or perturbation. 
Thus no energy is deposited irreversibly in the thermal part of the soft sector.
As shown below the hard sector only changes the way of how a pre-existing black hole is perceived at the boundary. Nevertheless, this amounts to a nontrivial back-reaction on the hard sector.

Moreover, since the simple situation presented below lies at the basis of the more complicated scenarios we want to address in the future, it is essential to study and to fully understand the dynamics of our test case.

\subsection{UV sector: Classical dynamics of homogeneous Yang-Mills fields }
\label{sec:YM}

The simplest configurations of the glasma, described by classical Yang-Mills equations, are produced by homogeneous color gauge fields. For simplicity, we shall consider SU(2) Yang-Mills theory.
Using temporal gauge $A_0^a=0$, with $a=1,2,3$, the Yang-Mills
equations of spatially homogeneous fields $A_i^a(t)$ are
a set of 9 coupled nonlinear ODE's. Setting $g=1$, they are given by
\begin{equation}\label{homYM}
 \ddot A_j^a{}-A_i^a A_i^b A_j^b+A_j^a A_i^b A_i^b=0,
\end{equation}
which immediately follows from $D^\mu F_{\mu\nu}=0$,
$F_{\mu\nu}^a=\6_\mu A_\nu^a-\6_\nu A_\mu^a+\epsilon^{abc}A_\mu^b A_\nu^c$,
and $\epsilon^{abc}\epsilon^{cde}=\delta^{ad}\delta^{be}-\delta^{ae}\delta^{bd}$.

In temporal gauge, Gauss' law, $D^\mu F_{\mu0}^d =0$, has to be imposed as a constraint. In
the homogeneous case it reduces to
\be\label{Gausslaw}
\epsilon^{dea}A^{ei}\dot A^a_i=0,
\ee
which is easy to satisfy by initial conditions where either
$A$ or $\dot A$ is set to zero.

The resulting dynamics of the 9 degrees of freedom 
is in general completely chaotic. The resulting energy-momentum
tensor $t_{\mu\nu}$ contains a conserved positive energy density, vanishing
Poynting vector $t_{0i}$, but the 
spatial stress tensor components (diagonal and nondiagonal) fluctuate with the only constraint of 4-dimensional tracelessness.

The energy-momentum tensor can be made diagonal by locking color and
spatial indices, i.e. assuming $A^a_i(t)\propto \delta^a_i$, which reduces the number of degrees of freedom from 9 to 3.
Switching on the semi-holographic coupling $\gamma$ in (\ref{effaction}) still allows us to restrict to a diagonal
tensor $\wT^{\mn}$, which gives extra source terms to the equations of motion (\ref{homYM}) that can be obtained explicitly from \eqref{eom}, but as can be explicitly checked the Gauss-law constraint (\ref{Gausslaw}) also remains unchanged.

The simplest nontrivial case is obtained by additionally requiring isotropy of the stress tensor.
Homogeneity and spherical symmetry with the same pressure in all three directions 
from fields with nontrivial time dependence  can be obtained by setting 
$A_1^1(t)=A_2^2(t)=A_3^3(t)=f(t)$.
Then the color electric fields are 
\be\label{YMfields}
E^a_i=\delta^a_i f',\quad B^a_i=\delta^a_i f^2,
\ee
forming a single anharmonic oscillator with
\be\label{oscillator}
 f''(t)+2f(t)^3=0.
\ee

\begin{figure}[t]
\begin{center}
\includegraphics[scale=1]{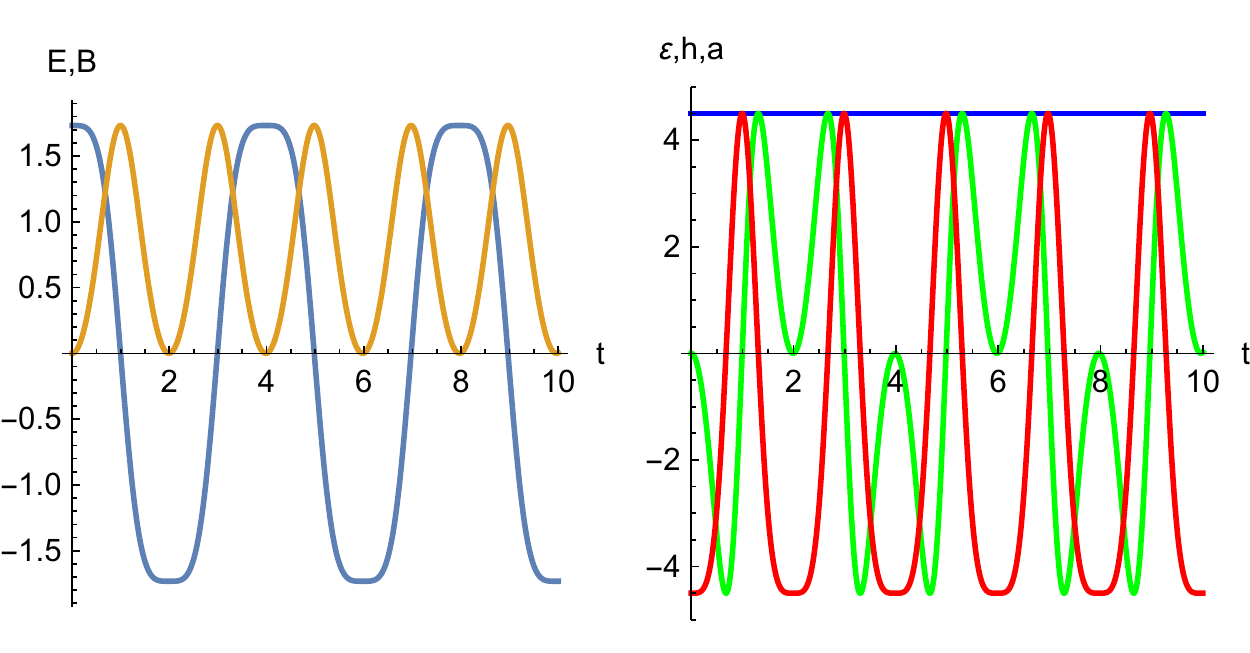}
\caption{Yang-Mills fields in our homogeneous and isotropic test case before
its coupling to an IR-CFT. Left: The electric (blue) and magnetic (orange) fields 
with $C=3^{1/4}$ and $t_0=0$ in (\ref{Jacobi}). Right: The conserved energy density $\varepsilon$ (blue)
and the oscillating quantities $h=
-\mathcal L_{\rm YM}$ (red) and $a=-\mathbf E^{a}\cdot\mathbf B^{a}$ (green).\label{fig:oscillator} }
\end{center}
\end{figure}

The expressions for the preserved energy and pressure read
\be\label{epspf}
\mathcal{\varepsilon}=3 p=\frac{1}{2}\(\mathbf E^{a2}+\mathbf B^{a2}\)=\frac{3}{2}\(f'(t)^2+f(t)^4\).
\ee
The equation of the anharmonic oscillator (\ref{oscillator}) has a closed-form solution in terms of the Jacobi elliptic function 
\be\label{Jacobi}
f(t)=C\,\mathrm{sn}(C(t-t_0)|-1)
\ee
which  is a double periodic function in the complex plane. Along the real
axis one has sinusoidal oscillations with the peculiarity that around a zero there is
a linear term but no cubic term because the potential is flat at the origin,
$\mathrm{sn}(t|-1)=t-t^5/10+O(t^9)$.

In the left panel of Fig.~\ref{fig:oscillator} the color-electric and color-magnetic fields are shown for initial
conditions corresponding to $C=3^{1/4}$ and $t_0=0$ in (\ref{Jacobi}). ($C$ is chosen such as to match
the numerical solutions of Fig.~\ref{fig:numericalsolution} in units where $Q_s=1$ and in the limit that the entire energy is in
the Yang-Mills fields.) The right panel shows the conserved energy density $\varepsilon$
and the oscillating quantities $h=-\frac12(\mathbf E^{a}\cdot\mathbf E^{a}-\mathbf B^{a}\cdot\mathbf B^{a})$ and $a=-\mathbf E^{a}\cdot\mathbf B^{a}$.
The negative action density $h$ is seen to oscillate between $-\varepsilon$ and $+\varepsilon$ without having a mean 
value of zero; the topological charge density $a$ has the same extremal values with a more complicated but
symmetric oscillation pattern.

When coupled to the IR-CFT, the energy-momentum tensor of the Yang-Mills theory is no longer conserved separately.
In order to use the above ansatz in a semi-holographic setup in the simplest way, we want to find  an appropriate exact gravity solution that allows for a  time dependent boundary metric of the form (\ref{gb}).

\subsection{IR-CFT sector: Analytic gravity solution}\label{sec:IR}

The simplest IR-CFT configuration that we can couple self-consistently to the classical Yang-Mills system considered above is one that  retains the same symmetries, namely homogeneity and isotropy. A remarkable simplification happens when we consider the case in which $\alpha$ and $\beta$ in (\ref{boundarycond}) vanish. 
The only non-trivial marginal deformation of the IR-CFT in this case is via the background metric (that is identified with the boundary metric of the dual gravity solution), which takes the form:
\begin{equation}
g^{\rm (b)}_{\mu\nu}=\eta_{\mu\nu}+\frac{\gamma}{Q_s^4}t_{\mu\nu}=\mathrm{diag}\left(-1+\frac{3\gamma}{Q_s^4}p(t),1+\frac{\gamma}{Q_s^4}p(t),1+\frac{\gamma}{Q_s^4}p(t),1+\frac{\gamma}{Q_s^4}p(t)\right).\label{eq:bc1}
\end{equation}
Note that in order to preserve the signature and regularity of the background metric the coupling $\gamma$ must lie in the interval:
\be
-\frac{Q_s^4}{p(t)}<\gamma<\frac{Q_s^4}{3p(t)}.
\ee
A crucial simplification for the following calculation is that
the metric (\ref{eq:bc1}) is \textit{conformally flat}.

As mentioned above, we impose homogeneity and isotropy on the IR-CFT energy-momentum tensor $\mathcal{T}^{\mu\nu}$. When the IR-CFT is holographic,  it follows that the dual geometry  should also have homogeneity and isotropy, since both the data, namely the boundary metric and $\mathcal{T}^{\mu\nu}$, which determine the solution, have these properties. Furthermore, as the hard-soft couplings $\alpha$ and $\beta$ are switched off, the bulk dilaton and axion fields vanish. This allows us to apply Birkhoff's theorem, according to which the bulk metric should be \textit{locally diffeomorphic} to the standard AdS-Schwarzschild black brane solution:
\begin{equation}
\mathrm{d}s^2=-\left(r^2-\frac{c}{r^2}\right)\mathrm{d}v^2+\left(r^2-\frac{c}{r^2}\right)^{-1}\mathrm{d}r^2+r^2\left(\mathrm{d}x^2+\mathrm{d}y^2+\mathrm{d}z^2\right),\label{eq:AdSSS}
\end{equation}
where $c$ is related to the mass of the black brane. In our case, $c$ will be a free parameter related to the IR-CFT initial conditions (particularly the initial energy in this sector).

The IR-CFT state dual to the AdS-Schwarzschild black brane (\ref{eq:AdSSS}) is the thermal state (with the temperature $T$ determined by the mass parameter $c$) living in flat Minkowski space, which is the boundary metric of this gravity solution. Since we are interested in a homogeneous and isotropic IR-CFT state that lives in the metric (\ref{eq:bc1}), we should apply
a bulk diffeomorphism which is non-trivial at the boundary and is such that the boundary metric transforms from $\eta_{\mu\nu}$ to (\ref{eq:bc1}). At the boundary, this bulk diffeomorphism should reduce to a precise combination of a Weyl transformation and a diffeomorphism of the time coordinate, since (\ref{eq:bc1}) is conformally flat. Indeed, such a bulk diffeomorphism is unique (once we also choose the final gauge), and is explicitly given in Appendix \ref{hol}. In Appendix \ref{hol} we also calculate the canonical charge of the bulk space time showing that up to anomalous effects the total gravitational energy is constant and equals that of the AdS-Schwarzschild black hole with a flat boundary. Applying this bulk diffeomorphism, we obtain our desired gravity solution from which we can obtain $\wT^{\mu\nu}$ defined in \eqref{eq:hatIRobjects} in the dual IR-CFT state, 
through the standard holographic dictionary that states:
\begin{equation}\label{eq:holoT}
\mathcal{T}^{\mu\nu}=\frac{2}{\sqrt{-g^{\rm{(b)}}}}\frac{\delta S_{\mathrm{EH}}^{\mathrm{on-shell}}}{\delta g^{\rm{(b)}}_{\mu\nu}},
\end{equation}
where $S_{\mathrm{EH}}^{\mathrm{on-shell}}$ denotes the holographically renormalized \textit{on-shell} five-dimensional Einstein-Hilbert action. Since $g^{\rm{(b)}}_{\mu\nu}$ in (\ref{eq:bc1}) is conformally flat, the energy-momentum tensor $\mathcal{T}^{\mu\nu}$ is not renormalization-scheme dependent.\footnote{Scheme dependence arises from the possibility of adding a covariantly conserved traceless tensor with the correct conformal weight to $\mathcal{T}^{\mu\nu}$ that is a local functional of the boundary metric (note that ${\rm Tr}\,\mathcal{T}$ evaluated in the background metric is scheme independent). In 4 dimensions this scheme-dependent contribution is proportional to the Bach tensor which vanishes on Weyl-flat (i.e. conformally flat) backgrounds.}

At this stage, it is interesting to note that we can also derive the form of the required homogeneous and isotropic $\mathcal{T}^{\mu\nu}$ in the IR-CFT via a general alternative method, which works also when the IR-CFT is not holographic. We still need to assume that, except for a non-trivial background metric, all other background sources vanish, which requires us to put the couplings $\alpha$ and $\beta$ to zero. The only additional assumption that we need to put in is that the IR-CFT state is the (time-dependent) conformal (plus a coordinate) transformation of the thermal state (with an arbitrary temperature), where the  transformation  brings $\eta_{\mu\nu}$ to the desired background metric (\ref{eq:bc1}) determined by the classical Yang-Mills fields. Since we cannot take the benefit of Birkhoff's theorem that applies on the gravity side in the large $N$ limit in holographic CFTs, we cannot say that this should be the unique homogeneous and isotropic IR-CFT living in this background metric. Nevertheless,
 
we 
can self-consistently assume that the IR-CFT state is the conformally transformed thermal state.

Under conformal transformation, $\mathcal{T}^{\mu\nu}$ transforms as a contravariant tensor of rank two and with Weyl weight two in a CFT, up to a \textit{state-independent} and renormalization-scheme independent anomalous term. In a 4D CFT, this anomalous term in a \textit{conformally flat} background metric is given by (see for example \cite{PhysRevD.16.1712,Herzog:2013ed}):
\be
\mathcal{T}^{{\rm (an)}\mu\nu} = -\frac{a_4}{(4\pi)^2}\left(g^{\mu\nu}\(\frac{R^2}{2}-R_{\alpha\beta}R^{\alpha\beta} \)+ 2R^{\mu\lambda}R^\nu_{\phantom{\mu}\lambda} - \frac{4}{3}RR^{\mu\nu}\right),
\ee
where $a_4$ is the central charge associated with the Euler density.\footnote{The central charge associated with the Weyl-tensor-squared term does not contribute here in a conformally flat background metric. As this anomalous piece is state-independent, we can obtain this from the vacuum in a conformally flat background metric.} Thus, we need to add the above anomalous piece (evaluated in the metric (\ref{eq:bc1}) to the covariant piece obtained after conformal transformation to compute the desired IR-CFT $\mathcal{T}^{\mu\nu}$. In a holographic CFT, as expected on general grounds, we get the same result in this method as obtained via the direct application of the holographic dictionary on the dual gravity solution as mentioned before, by using
\be
a_4 = \frac{N_c^2}{4}.
\ee
The above is a universal result for a strongly coupled large $N$ holographic CFT.

Either by using the standard holographic dictionary (for details see Appendix \ref{hol})
or via the simplified procedure mentioned above, we obtain the (diagonal) components of $\wT^{\mu\nu}$ as follows:
\begin{eqnarray}\label{eq:emt}
\wE &:=&\wT^{tt}=\frac{N_c^2}{2\pi^2}\left(\frac{3 c}{4 r_{\ms{(0)}}(t)^2 v_{\ms{(0)}}'(t)}+\frac{3 r_{\ms{(0)}}'(t)^4}{16 r_{\ms{(0)}}(t)^6 v_{\ms{(0)}}'(t)^5}\right),\nonumber\\
\wP &:=&\wT^{xx}=\wT^{yy}=\wT^{zz}=\nonumber\\
&=&\frac{N_c^2}{2\pi^2}\left\{\frac{c v_{\ms{(0)}}'(t)}{4 r_{\ms{(0)}}(t)^2}+\frac{r_{\ms{(0)}}'(t)^2 \left[4 r_{\ms{(0)}}(t) r_{\ms{(0)}}'(t) v_{\ms{(0)}}''(t)+r_{\ms{(0)}}(t) \left(5 r_{\ms{(0)}}'(t)^2-4 r_{\ms{(0)}}(t) r_{\ms{(0)}}''(t)\right)\right]}{16 r_{\ms{(0)}}(t)^6 v_{\ms{(0)}}'(t)^4}\right\},\nonumber\\
\end{eqnarray}
with
\begin{eqnarray}\label{coeff2}
r_{\ms{(0)}}(t)=\sqrt{1+(\gamma/Q_s^4)p(t)}~,\nonumber\\
v_{\ms{(0)}}'(t)=\sqrt{\frac{1-(\gamma/Q_s^4)3p(t)}{1+(\gamma/Q_s^4)p(t)}}~.
\end{eqnarray}
The terms proportional to $c$ are simply the result of the transformation of the energy-momentum tensor density of the AdS-Schwarzschild space-time with flat boundary conditions and thus only involve the Yang-Mills pressure $p(t)$ without derivatives. This part of the energy-momentum tensor is still traceless and (covariantly) conserved on its own
with respect to the metric $g^{\rm (b)}_{\mn}$. Furthermore, it is also the most relevant contribution to the energy-momentum density in a small $\gamma/Q_s^4$ expansion which reads
\begin{eqnarray}
\wE &=&\frac{3cN_c^2}{8\pi^2}\left[1+\frac{\gamma}{Q_s^4}p(t)+3\left(\frac{\gamma}{Q_s^4}p(t)\right)^2\right]+\mathcal{O}\left[\left(\frac{\gamma}{Q_s^4}p(t)\right)^3\right]~,\label{eq:IRenergy}\\
\wP &=&\frac{cN_c^2}{8\pi^2}\left[1-3\frac{\gamma}{Q_s^4}p(t)+3\left(\frac{\gamma}{Q_s^4}p(t)\right)^2\right]+\mathcal{O}\left[\left(\frac{\gamma}{Q_s^4}p(t)\right)^3\right].\label{eq:IRpressure}
\end{eqnarray}
As we shall see below, only the combination $(\gamma/Q_s^4)(\wE +\wP )$ contributes
in the equations of motion of the semi-holographic model.
Therefore the leading-order effect of the back-reaction on the Yang-Mills fields, mediated by the deformation of the IR-CFT due to the Yang-Mills fields which determine $p(t)$, is of order $(\gamma/Q_s^4)^3$. To leading order $\gamma/Q_s^4$, the IR-CFT acts on the Yang-Mills fields through the soft thermal bath that one starts with by choosing $c\neq0$.

The terms independent of $c$ in Eq. \eqref{eq:emt} are due to the nonvanishing Ricci tensor of $g^{\ms{(b)}}_{\mu\nu}$ and their trace gives the conformal anomaly, which in our case is determined solely by the Euler-density associated with $g^{\ms{(b)}}_{\mu\nu}$. These only contribute at non-stationary points of the Yang-Mills pressure, i.e., for $p'(t)\neq 0$, and are of order $(\gamma/Q_s^4)^3$.

\subsection{Coupling the UV with the IR sector}
We now have all the ingredients at hand to test our modified  semi-holography proposal. 
The  equations of motion (\ref{eom}) for the coupled system  with a homogeneous and isotropic ansatz  for the UV-Yang-Mills gauge fields  (\ref{YMfields}) coupled to a homogeneous and isotropic IR energy-momentum tensor  become
\begin{equation}
f''(t)+2  \frac{1-\frac{1}{2} \frac{\gamma}{Q_s^4}(\wE +\wP )}{1+\frac{1}{2} \frac{\gamma}{Q_s^4}(\wE +\wP )}f(t)^3+\frac{1}{2}  \frac{\gamma}{Q_s^4}\frac{(\wE +\wP )'}{1+\frac{1}{2} \frac{\gamma}{Q_s^4}(\wE +\wP )} f'(t)=0,\label{eq:EOMiso}
\end{equation}
where $\wE $ and $\wP $ are given by (\ref{eq:emt}). It is  remarkable that  only the combination
$\wE+\wP$ is needed, which  is given in closed form by
\begin{eqnarray}\label{EpP}
\wE +\wP &=&\frac{N_c^2}{2\pi^2}\frac{c}{\sqrt{1-3 \gb   p(t)}[1+\gb  p(t)]^{3/2}}\nonumber\\
&\ &+\frac{N_c^2}{2\pi^2}\frac{\gb ^3 p'(t)^2 \left(2 [1+\gb  p(t)] [3 \gb   p(t)-1] p''(t)-\gb   [1+6\gb   p(t)] p'(t)^2\right)}{64 [1-3\gb   p(t)]^{5/2} [1+\gb p(t)]^{7/2}}\;,
\end{eqnarray}
where we introduced the abbreviation $\gb =\gamma/Q_s^4$. 

The full energy density for our test case is obtained by taking the zeroth component of the energy-momentum tensor (\ref{EMT}),
which can be expressed  in the following convenient form:
\begin{equation}
E=\varepsilon+\wE \left(1- \frac{\gamma}{Q_s^4}  \varepsilon\right)+\frac{3}{2}\frac{\gamma}{Q_s^4}\left(\wE +\wP \right)f'(t)^2.
\end{equation}

In order to solve \eqref{eq:EOMiso}, which is an ordinary fourth-order differential equation when $p(t)$ in $\wE +\wP$
is expressed in terms of $f(t)$ through \eqref{epspf}, we will use the iterative algorithm proposed in \cite{Iancu:2014ava} and test its usability.
By interpreting $\wE +\wP $ as a fixed external source,  which is updated after each iteration, this reduces to a second-order differential equation for  $f(t)$.
While in principle this method should not be necessary for our test case, 
in practice we have not been able to find regular solutions of the above highly nonlinear fourth-order equation other than by following the iterative procedure. Moreover,
one has to keep in mind that for more complicated dynamics, e.g., for an anisotropic situation or for $\alpha,\beta\neq 0$, the expression of $\wT^{\mu\nu}$ in terms of the Yang-Mills fields
(here $f(t)$) will not be explicitly known.

Eq. \eqref{eq:EOMiso} is reminiscent of an equation of motion for an anharmonic oscillator where the coefficient of the $f(t)^3$ term determines the frequency of the oscillations, whereas the last term acts as (anti-)damping term. In order to obtain regular oscillating solutions the frequencies are constrained to take real values. 

Before discussing the full numerical result, we describe  the first two steps of the algorithm,  allowing  us to extract  analytic behavior which approximates   the full solution very well for sufficiently small values of  $N_c^2\gamma c/(2\pi^2Q_s^4)$.
In the first step we set $\wE +\wP =0$ which amounts to the solution for $f(t)$ we have already discussed in section \ref{sec:YM} and for which $p(t)=p_{\ms{0}}:=p(0)$. Inserting the latter into \eqref{EpP}, the updated source now reads
\begin{equation}
\wE _{\ms{0}}+\wP _{\ms{0}}=\frac{N_c^2}{2\pi^2}\frac{c}{\sqrt{1-3 \frac{\gamma}{Q_s^4}  p_{\ms{0}}}[1+\frac{\gamma}{Q_s^4}  p_{\ms{0}}]^{3/2}},\label{eq:initialEplusP}
\end{equation}
which is larger than $N_c^2/(2\pi^2)c$ irrespective of the sign of $\gamma$.

For a special choice of initial conditions, either $f'(0)=\pm\sqrt{2p_{\ms{0}}}$, $f(0)=0$ or $f'(0)=0$, $f(0)=\pm(2p_{\ms{0}})^{1/4}$, it follows automatically that $\wE +\wP $ at the initial time $t=0$ always takes the value given in \eqref{eq:initialEplusP}.\footnote{The case $f'(0)=0$, $f(0)=\pm(2p_{\ms{0}})^{1/4}$ implies $p'(0)=0$ while in the case $f'(0)=\pm\sqrt{2p_{\ms{0}}}$, $f(0)=0$ one has $(\wE +\wP )'(0)\propto p'(0)=f'(0)f''(0)$ and hence one obtains $f''(0)=0$ from the equations of motion.}

\begin{figure}[t]
\centerline{
\includegraphics[scale=0.7]{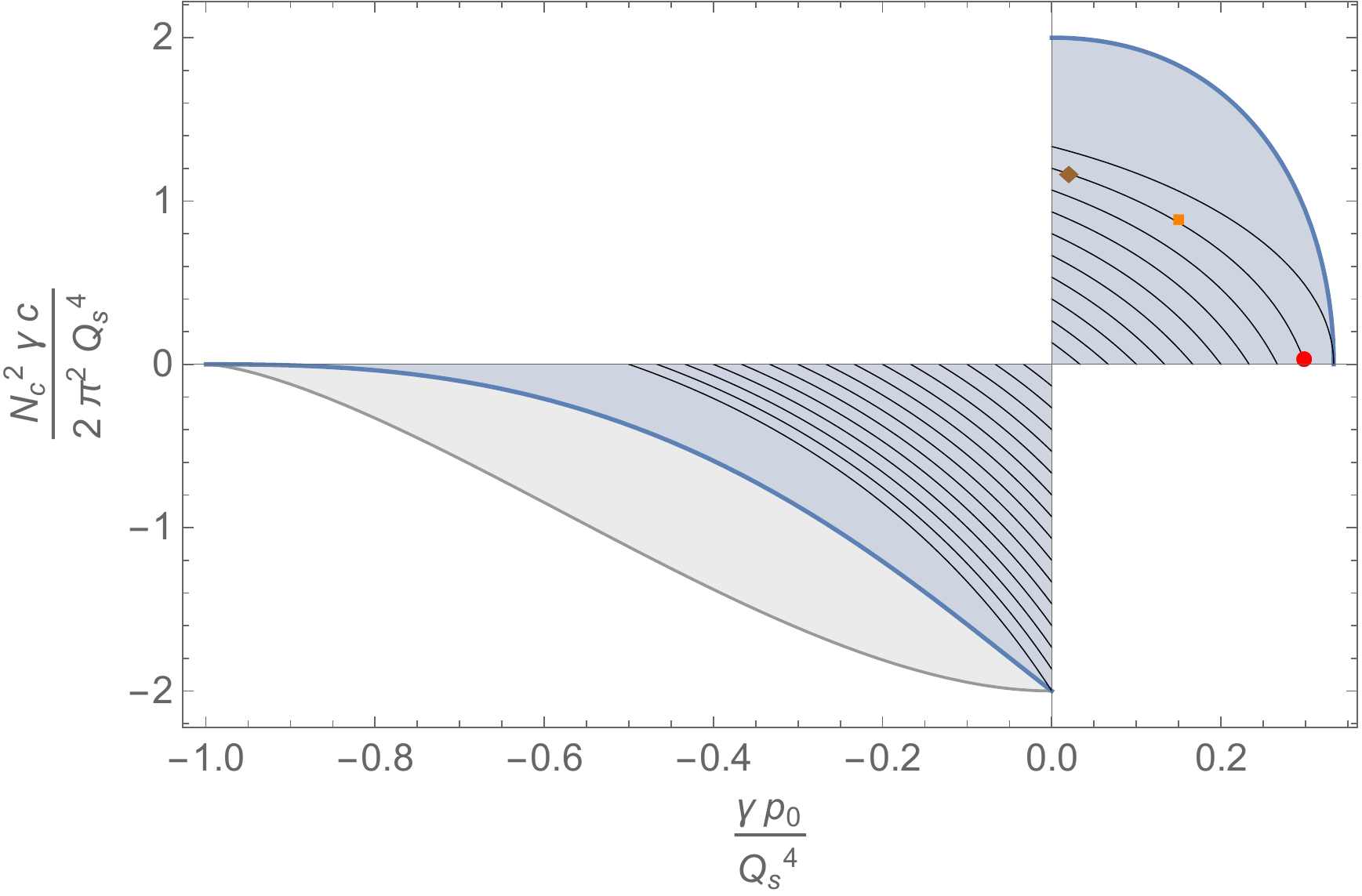}}
\caption{The allowed regions (shaded) for the values of $\frac{N_c^2\gamma c}{2\pi^2Q_s^4}$ and $\frac{\gamma p_{\ms{0}}}{Q_s^4}$. In the lower left quadrant the lower boundary curve (grey) is obtained from the reality condition for $\omega$ in \eqref{eq:frequency}, i.e. after the first iteration. The more restrictive boundary curve (blue) above is obtained from demanding the regularity of the metric after the second iteration step. The thin lines correspond to constant values of the total energy $\frac{\gamma E}{Q_s^4}=i/10$, with $i=1,\dots 10$ in the first quadrant and $i=-1,\dots -15$ in the third quadrant. The point, the square and the diamond correspond to the values used for the numerical evaluations.\label{fig:gammabounds}}
\end{figure}

In the second step of the iterative process the derivative of $\wE +\wP $ still vanishes and   we only have to solve the second order differential equation
\begin{equation}
f''(t)+2 \frac{1-\frac{1}{2} \frac{\gamma}{Q_s^4}(\wE _{\ms{0}}+\wP _{\ms{0}})}{1+\frac{1}{2} \frac{\gamma}{Q_s^4}(\wE _{\ms{0}}+\wP _{\ms{0}})}f(t)^3 =0\;.
\end{equation}
Choosing  the initial conditions $f(0)=(2p_{\ms{0}})^{(1/4)}$ and  $f'(0)=0$, again   an  analytic  solution  in terms of Jacobi elliptic functions can be found,
\begin{eqnarray}
f(t)&=&(2p_{\ms{0}})^{\frac{1}{4}}\mathrm{cd}(\omega t\vert-1)\;,\nonumber\\
\omega&:=&4K(-1)\nu=(2p_{\ms{0}})^{\frac{1}{4}}\left(\frac{1-\frac{1}{2} \frac{\gamma}{Q_s^4}(\wE _{\ms{0}}+\wP _{\ms{0}})}{1+\frac{1}{2} \frac{\gamma}{Q_s^4}(\wE _{\ms{0}}+\wP _{\ms{0}})}\right)^{\frac{1}{2}}\;,\label{eq:frequency}
\end{eqnarray}
where $K(-1)\approx 1.31$ denotes the complete elliptic integral of the first kind, which is the quarter period of the Jacobi elliptic function, e.g. $\mathrm{sn}(x\vert-1)=\mathrm{sn}\left(x+4K(-1)\vert-1\right)$. Accordingly we may call $\nu$ the frequency of the anharmonic oscillations. As already mentioned, demanding a regular solution imposes a reality condition on $\nu$, which in turn puts bounds to the value of $N_c^2\gamma c/(2\pi^2Q_s^4)$ in addition to those put on $\gamma/Q_s^4 p_{\ms{0}}$ guaranteeing regularity of $g^{\ms{(b)}}_{\mu\nu}$. The allowed parameter space  is illustrated in Fig.~\ref{fig:gammabounds}.
The total initial energy of the full semi-holographic system for this choice of initial conditions is given by
\begin{eqnarray}
E&=&\varepsilon(0)+\wE (0)\left(1-\frac{\gamma}{Q_s^4}\varepsilon(0)\right)=3p_{\ms{0}}+\frac{3N_c^2c}{8\pi^2}\sqrt{\frac{1-3\frac{\gamma}{Q_s^4}p_{\ms{0}}}{1+\frac{\gamma}{Q_s^4}p_{\ms{0}}}},\label{eq:totalenergy}
\end{eqnarray}
which remains true to all orders of the iterative algorithm. 

\begin{figure}[h]
\begin{center}
\includegraphics[scale=0.7]{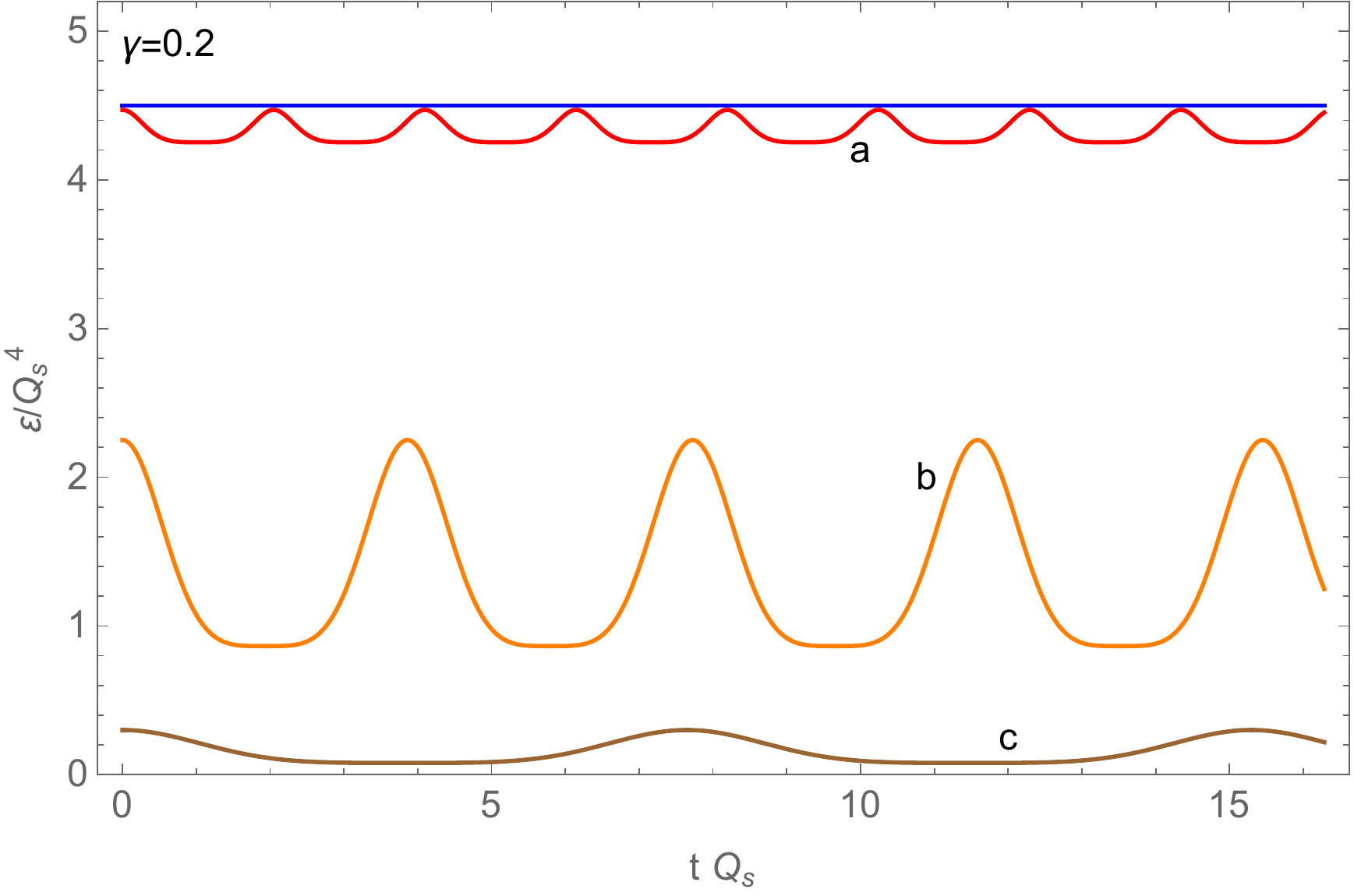}
\qquad\includegraphics[scale=0.7]{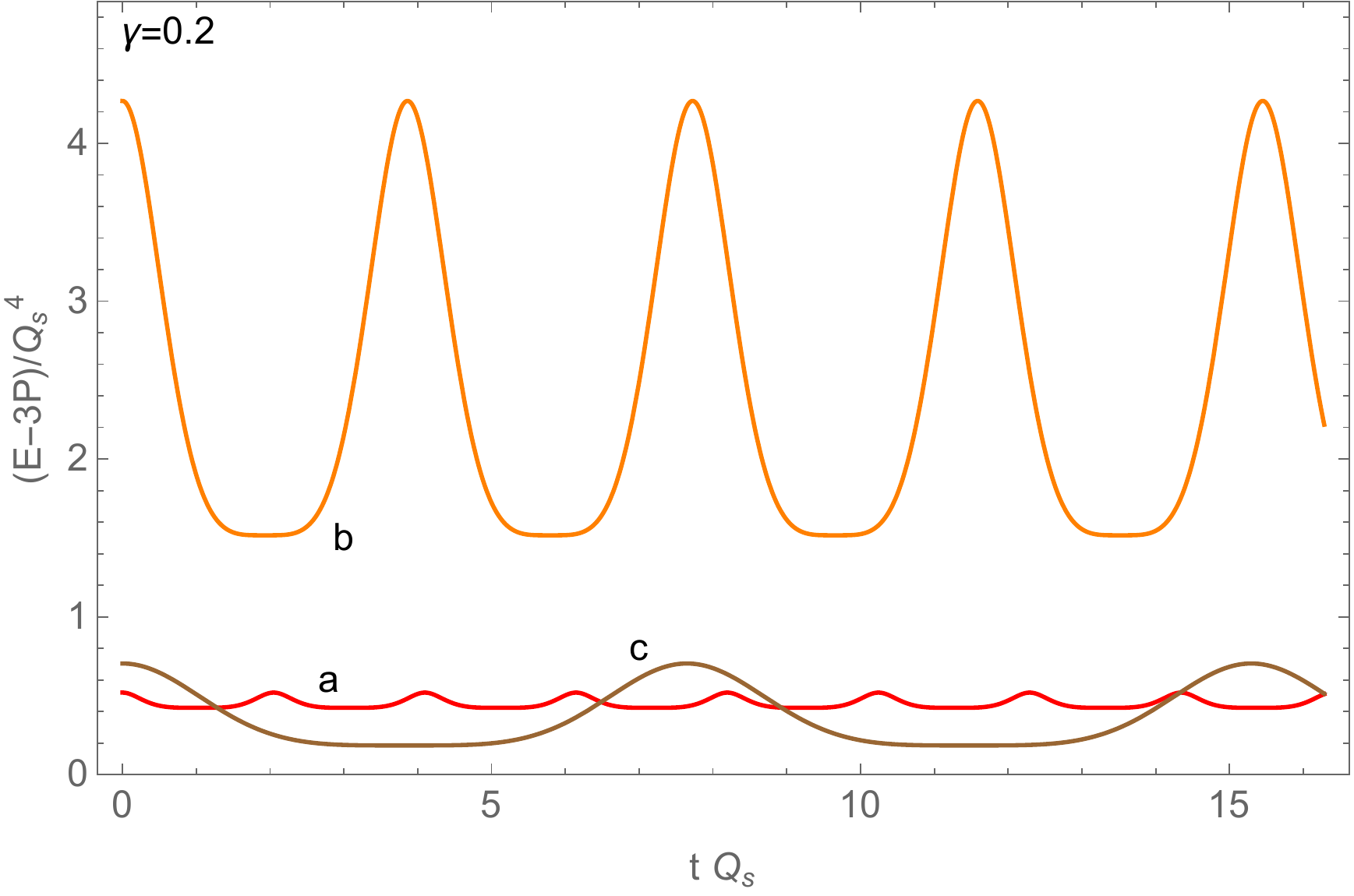}
\caption{The numerical solution of \eqref{eq:EOMiso} in terms of the Yang-Mills energy density
$\varepsilon$ (upper panel) and
the negative trace of the full energy-momentum tensor, $-T^{\mn}\eta_{\mn}=E-3P$, (lower panel) for $p^{\ms{(a)}}_{\ms{0}}/Q_s^4=1.49$, $p^{\ms{(b)}}_{\ms{0}}/Q_s^4=0.75$, $p^{\ms{(c)}}_{\ms{0}}/Q_s^4=0.1$ with $\gamma=0.2$ and total (conserved) energy density $E/Q_s^4=4.5$, marked by the blue line in the upper panel.
The three solutions (a), (b), (c) involve increasing values of $c$,
and $\varepsilon$ constitutes a correspondingly smaller fraction
of the total energy density $E$ (upper panel). The trace term $E-3P$ in the lower panel can be
thought of as an ``interaction measure'' of the UV and IR sector which in our test case
vanishes exactly in the two extreme cases $\varepsilon=E$ and $\varepsilon=0$.\label{fig:numericalsolution}
}
\end{center}
\bigskip
\end{figure}

Starting with the third step of the iterative algorithm requires to solve the second order equation for $f(t)$ numerically. Note that for positive $\gamma$ we find $p(t)\leqslant p_{\ms{0}}$, while for negative $\gamma$ we find  $p(t)\geqslant p_{\ms{0}}$, meaning that for negative $\gamma$ the allowed region for the values of $N_c^2\gamma c/(2\pi^2Q_s^4)$ is further restricted in order to ensure  regularity of the metric as shown by the blue line in Fig.~\ref{fig:gammabounds}. 
There we also show lines of constant values of $\gamma E/Q_s^4$ that completely fit within the restricted regions. For positive $\gamma$ we find $\gamma E/Q_s^4\leqslant 1$ and for negative $\gamma$ this condition amounts to $\gamma E/Q_s^4\geqslant -1.5$.

In Fig.\ \ref{fig:numericalsolution} we present the numerical solution for three different choices of $p_{\ms{0}}/Q_s^4$ with $\gamma=0.2$ and constant energy $E/Q_s^4=4.5$: $p^{\ms{(a)}}_{\ms{0}}/Q_s^4=1.49$, $p^{\ms{(b)}}_{\ms{0}}/Q_s^4=0.75$, $p^{\ms{(c)}}_{\ms{0}}/Q_s^4=0.1$. These choices are depicted by the colored markers in Fig.~\ref{fig:gammabounds}.

\begin{figure}[t]
\begin{center}
\includegraphics[scale=0.7]{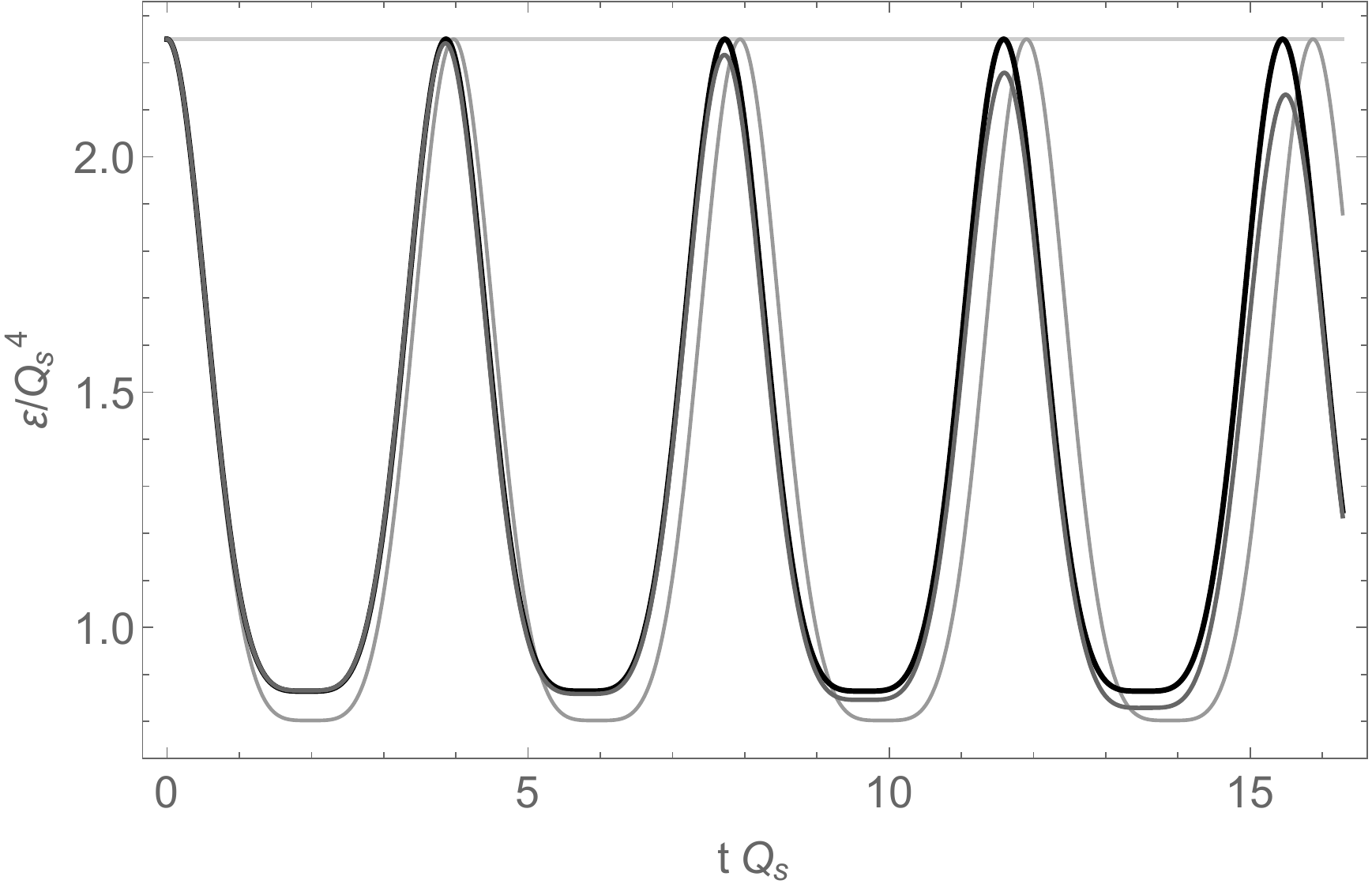}
\caption{Convergence of the iterative solution to case (b) of Fig.\ \ref{fig:numericalsolution}, with increasing darkness of the grey lines for increasing order of the iteration. From the fourth iteration onwards the changes are too small to be visible in this plot.\label{fig:convergence}
}
\end{center}
\end{figure}

In the upper panel of Fig.\ \ref{fig:numericalsolution}, 
the blue horizontal line
marks the total energy density, and the three oscillating lines give the 
Yang-Mills part of it for the three choices of $p_{\ms{0}}$. The difference
to the total energy density is the oppositely oscillating component contained in the
soft modes together with their interaction energy with the hard modes.\footnote{Although 
the entropy of the dual black hole and its intrinsic temperature
is unchanging, the changing boundary metric $g^{({\rm b})}_{\mn}$ leads to a changing 
energy density contribution $\wT^{00}$ in (\ref{EMT}), as further discussed in Appendix B.
However, as we discussed above, there is no unambiguous separation of purely soft
contributions from the explicit soft-hard interaction terms in the full energy-momentum tensor (\ref{EMT}).
E.g., rewriting everything in terms of $\mathcal{T}^{\mn}$ instead of $\wT^{\mn}$ would
suggest a different split of the contributions coming from $W_{\mathrm{CFT}}[g^{({\rm b})}_{\mn}]$
into soft and soft-hard parts, again with both changing in time. 
From the constancy of the entropy of the black hole one can however
conclude that there is no change in the {\it thermalized part} of the energy in the soft sector.}

Fixing the energy density in (\ref{eq:totalenergy}) and varying $p_{\ms{0}}/Q_s^4$ 
for the three solutions
means that we have to adjust the parameter $c$, which is related to the black holes mass, accordingly. 
Going from higher to lower values in $p_{\ms{0}}$ with fixed $E$ means that $c$ is increased and the black hole made larger. 
(The case $c=0$ corresponds to the trivial solution where one has exactly $p_{\ms{0}}=E/3$.)
In a more realistic setup where the gravity solution is dynamical and not just a gauge transformation of the AdS-Schwarzschild black hole, we expect that due to the interaction between the hard and soft sector the growing black hole draws energy from the hard modes as the system thermalizes. The UV energy density thus may shrink and get deposited in the IR-sector. This 
is mimicked here by choosing different initial conditions. 
Interestingly, with the choice of positive $\gamma$ in the example shown in Fig.~\ref{fig:numericalsolution}, 
the wavelength of the oscillation increases as the Yang-Mills energy density decreases, which is not the case for negative $\gamma$.
Furthermore, for positive $\gamma$
the trace term $E-3P$ depicted in the lower panel of Fig.~\ref{fig:numericalsolution} is always positive, consistent with lattice results of QCD thermodynamics \cite{Boyd:1996bx,Borsanyi:2010cj}, where $E-3P$ is often called the ``interaction measure''.

We applied  two criteria  in order to decide whether the algorithm converges. The first is of course to check whether equation \eqref{eq:EOMiso} evaluated at the solution for $f(t)$ is satisfied sufficiently, which is nontrivial, since the solution for $f(t)$ was obtained for source terms  evaluated with the  solution of the previous iteration step. The second criterion is that the total energy evaluated at the solution for $f(t)$ is sufficiently close to the value prescribed in Eq. \eqref{eq:totalenergy} respectively. In the solution presented in 
Fig.~\ref{fig:numericalsolution} both criteria were satisfied up to order $10^{-7}$ which is smaller than any term appearing in \eqref{eq:EOMiso} in particular the one involving the fourth order derivative of $f(t)$.

Figure \ref{fig:convergence} shows the convergence of the iterative solution for the case (b). From the fourth iteration onwards, there is no visible change of the numerical result.

\section{Conclusions}\label{sec:conclusions}

In this paper we presented an improved and extended version of the semi-holographic model proposed in \cite{Iancu:2014ava}.
The idea behind semi-holography is to construct a phenomenological model for heavy-ion collisions  that is able to incorporate the interaction between the weakly coupled UV and the strongly coupled IR-sector. 
This is done by  assuming  that the UV sector is given by over-occupied gluon modes
admitting a description in terms of classical Yang-Mills fields, while the strongly coupled
IR-sector is described by a conformal field theory with a holographic dual. 

We presented a new action where such a mechanism can be implemented, and we constructed the full energy-momentum tensor of the coupled theory which is locally conserved and thus can track the energy transfer between the two sectors. The full energy-momentum tensor consists of the standard Yang-Mills theory expression plus contributions derived from the effective action of
the conformal field theory which is deformed by the marginal operators of the Yang-Mills theory. The latter contributions thus
comprise both purely soft and soft-hard ones, which however cannot be disentangled unambiguously; only the thermal part of
the soft sector has a well-defined meaning in terms of the black hole horizon of the holographic gravity dual.
 
Moreover, we have presented first numerical tests of the iterative procedure proposed for solving the evolution
of the semi-holographic model.
By  assuming homogeneity and isotropy we were able to construct a simple test case with closed analytic solutions for the two decoupled sectors. Coupling them together via the tensorial coupling only, we numerically solved for the evolution of the fields and observed quick convergence of the algorithm and observed
energy exchange between the Yang-Mills part and the conformal field theory with constant total energy.
This is already a nontrivial result and is a proof of principle of this semi-holography proposal. 

Of course, due to the simplicity and high symmetry of this test case we could not observe thermalization. 
This was expected because without any propagating degrees of freedom our gravity solution does not incorporate  dynamical black hole formation,
only time-dependent diffeomorphisms of a pre-existing black hole, i.e., the black hole temperature and the entropy measured by the surface gravity and the horizon area respectively are not altered.
In order for the system to relax, more degrees of freedom must be included.
This can be done by introducing  anisotropy in the space-time along the lines of \cite{Chesler:2008hg}  and also making the Yang-Mills fields anisotropic. It is then expected that the black hole is able to draw energy from the hard modes and that isotropization occurs. 
Another possibility is to stay isotropic and to turn on one or both of the scalar fields
 on the gravity side coupled to the Lagrange density and the Pontryagin density on the field theory side, respectively. 
The latter would be particularly interesting in view of anomalous transport phenomena such as the chiral
magnetic effect \cite{Kharzeev:2015znc} and also with regard to the results of Ref.~\cite{Rebhan:2011vd}
on the shear viscosity in strongly coupled anisotropic systems.
We plan to tackle these more complicated scenarios in future work.

Depending on the initial conditions and/or the values of $\alpha$, $\beta$ and $\gamma$, various different scenarios of thermalization may appear, e.g.,
the hard modes may eventually drain their energy almost completely to the black hole, or
the hard modes might retain a significant fraction of their energy even at late time, while the full energy-momentum tensor (\ref{EMT}) approaches a diagonal form.
In the latter case, we would expect that  the effective hydrodynamic description which results from the fluid/gravity correspondence \cite{Rangamani:2009xk} will not suffice to describe the late stage of the evolution prior to hadronization. Since at late times, where the relatively hard gluons are less densely populated,
a kinetic theory approach may be more suitable than 
the classical field theory (glasma) description, it would also be
interesting to understand  how to couple kinetic theory to a holographic CFT
in our semi-holographic framework. 
In this context, it may be useful to follow
\cite{Muller:2015maa,Yang:2015bva}, where a similar attempt has been made in context of constructing a
kinetic theory of leptons and photons when the latter are interacting with
strongly coupled QCD matter. This could e.g.\ have an important bearing on the $v_2$ photon puzzle \cite{Adare:2011zr,Lohner:2012ct} that pertains to the late stage of the evolution.

In the future, we may also hope to determine the (so far free) hard-soft coupling constants $\alpha$, $\beta$, and $\gamma$ in terms of $\alpha_s(Q_s)$ to a certain degree of approximation, by using renormalisation group techniques as discussed in \cite{Behr:2015yna,Behr:2015aat}. In this case, the semi-holographic model could eventually lead to sharp predictions.

\section*{Acknowledgments}

We thank C.\ Ecker, N.\ Gaddam, E.\ Iancu, and D.-L. Yang for useful discussions. 
A.\ Mukhopadhyay acknowledges support from a Lise Meitner fellowship
of the Austrian Science Fund (FWF), project no.\ M1893-N27;
F.\ Preis and S.\ Stricker were supported by the FWF project P26328-N27.

\begin{appendix}
\section{Conservation of the full energy-momentum tensor}\label{EMTconserv}
In this Appendix we show that the  full energy-momentum tensor
\begin{eqnarray}\label{EMTapp}
T^{\mu\nu}&=&t^{\mu\nu}\nonumber\\
&\ &+\wT^{\alpha\beta}\left\{\delta_{(\alpha}^\mu\delta_{\beta)}^\nu-\frac{\gamma}{Q_s^4N_c}\left[\mathrm{Tr}(F_\alpha^{\ \mu}F_\beta^{\ \nu})-\frac{1}{2}\eta_{\alpha\beta}\mathrm{Tr}(F^{\mu\rho}F^{\nu}_{\ \rho})+\frac{1}{4}\delta_{(\alpha}^\mu\delta_{\beta)}^\nu\mathrm{Tr}(F^2)\right]\right\}\nonumber\\
&\ &-\frac{\beta}{Q_s^4N_c}\wH\,\mathrm{Tr}(F^{\mu\alpha}F^{\nu}_{\ \alpha})-\frac{\alpha}{Q_s^4}\eta^{\mu\nu}\wA\, a
\end{eqnarray}
is conserved to all orders in the coupling constants. 
Let us first revisit the divergence of $t^{\mu\nu}$ for completeness:
\begin{equation}
\partial_\mu t^{\mu\nu}=\frac{1}{N_c} \mathrm{Tr}[(D_\mu F^{\mu\rho})F^{\nu}_{\ \rho}]
\end{equation}
where we used
\begin{equation}
\frac{1}{N_c}\mathrm{Tr}(F^{\mu\rho}D_\mu F^{\nu}_{\ \rho}-\frac{1}{2}F^{\rho\mu}D^\nu F_{\rho\mu})
=-\frac{1}{2N_c}\mathrm{Tr}[F^{\rho\mu}\eta^{\nu\tau}(D_\tau F_{\rho\mu}+D_\rho F_{\mu\tau}+D_\mu F_{\tau\rho})]=0,\qquad
\end{equation}
due to the Bianchi identity.
In order to show conservation of the full energy-momentum tensor we need to rewrite the individual interaction terms. 
The first term can be brought  in the following form
\begin{eqnarray}\label{first}
&&-\frac{\gamma}{Q_s^4N_c}\partial_\mu[\wT^{\alpha\beta}\mathrm{Tr}(F_\alpha^{\ \mu}F_\beta^{\ \nu})]\nonumber\\
&&=-\frac{\gamma}{Q_s^4N_c}\mathrm{Tr}[D_\mu(F_{\ \alpha}^{\mu}\wT^{\alpha\beta})F^{\nu}_{\ \beta} ]-\frac{\gamma}{Q_s^4N_c}\mathrm{Tr}(\wT^{\alpha\beta}F_\alpha^{\ \mu}D_\mu F_\beta^{\ \nu} ),
\end{eqnarray}
where the first term on the right-hand side appears in the equations of motion \eqref{eom}.
By using the same trick as for the YM energy-momentum tensor the second term becomes
\begin{eqnarray}\label{second}
&&\frac{\gamma}{2Q_s^4N_c}\partial_\mu[\wT^{\alpha\beta}\eta_{\alpha\beta}\mathrm{Tr}(F^{\mu\rho}F^{\nu}_{\ \rho})]\nonumber\\
&&=\frac{\gamma}{2Q_s^4N_c}\mathrm{Tr}[D_\mu(\wT F^{\mu\beta})F^{\nu}_{\ \beta}]+\frac{\gamma}{8Q_s^4N_c}\wT\partial^\nu \mathrm{Tr}(F^2),
\end{eqnarray}
where  also the first term appears in the equations of motion.
The third term is
\begin{eqnarray}\label{third}
-\frac{\gamma}{4Q_s^4N_c}\partial_\mu[\wT^{\mu\nu}\mathrm{Tr}(F^2)]=\frac{\gamma}{Q_s^4}\partial_\mu\{\wT^{\mu\alpha}[t_\alpha^{\ \nu}-\frac{1}{N_c}\mathrm{Tr}(F_\alpha^{\ \beta}F^{\nu}_{\ \beta})]\} \qquad\nonumber\\
=-\frac{\gamma}{Q_s^4N_c}\mathrm{Tr}[D_\mu(\wT^{\mu\alpha}F_\alpha^{\ \beta})F^{\nu}_{\ \beta}]+\frac{\gamma}{Q_s^4}\partial_\mu(\wT^{\mu\alpha}t_\alpha^{\ \nu})-\frac{\gamma}{2Q_s^4}\wT^{\alpha\beta}\partial^\nu t_{\alpha\beta}\nonumber\\
-\frac{\gamma}{8Q_s^4N_c}\wT\partial^\nu \mathrm{Tr}(F^2)+\frac{\gamma}{Q_s^4N_c}\mathrm{Tr}(\wT^{\alpha\beta}F_\alpha^{\ \mu}D_\mu F_\beta^{\ \nu} ),
\end{eqnarray}
where  the Bianchi identity was used to obtain the last two terms, which cancel the unwanted terms from (\ref{first}) and (\ref{second}).
Again, the first term on the right hand side appears in the equations of motion.

The fourth term  involves the dilaton field and is 
\begin{eqnarray}
-\frac{\beta}{Q_s^4N_c}\mathrm{Tr}[D_\mu(F^{\mu\beta}\wH F^{\nu}_{\ \beta})]=-\frac{\beta}{Q_s^4N_c}\mathrm{Tr}[D_\mu(F^{\mu\beta}\wH)F^{\nu}_{\ \beta}]-\frac{\beta}{Q_s^4}\partial^\nu h\wH,
\end{eqnarray}
where again the first term appears in the equations of motion. Similarly, the last term reads
\begin{eqnarray}
-\frac{\alpha}{Q_s^4}\partial^\nu \left(a\wA\right)=-\frac{\alpha}{N_cQ_s^4}\partial_\mu\wA\mathrm{tr}(\tilde{F}^{\mu\beta}F^{\nu}_{\ \beta})-\frac{\alpha}{Q_s^4}\partial^\nu a\wA,
\end{eqnarray}
where we used the identity\footnote{The identity (\ref{eq:CSid}) is valid for an arbitrary antisymmetric tensor
$F_{\mn}$ and its dual in four dimensions.
It can also be confirmed by considering the expression for the electromagnetic energy-momentum tensor in
a linear medium, which reads $D^{\mu\beta}F^{\nu}_{\ \beta}-\frac14\eta^{\mn}D^{\alpha\beta}F_{\alpha\beta}$, where $D^{\mn}$
involves the fields $\mathbf D$ and $\mathbf H$ in place of $\mathbf E$ and $\mathbf B$.
The familiar components of this tensor are the energy density $\epsilon=(\mathbf D\cdot\mathbf E+\mathbf H\cdot \mathbf B)/2$,
the momentum density $\mathbf D\times\mathbf B$, the Poynting vector $\mathbf E\times\mathbf H$, and the stress tensor
$D_i E_j+B_i H_j-\delta_{ij}\epsilon$. All those vanish identically when one sets $D^{\mn}=\tilde F^{\mn}$, i.e., when $\mathbf D=\mathbf B$
and $\mathbf H=-\mathbf E$.}
\begin{equation}
\frac{1}{N_c}\mathrm{tr}(\tilde{F}^{\mu\beta}F^{\nu}_{\ \beta})=\frac{1}{4N_c}\eta^{\mu\nu}\mathrm{tr}(\tilde{F}^{\alpha\beta}F_{\alpha\beta})=\eta^{\mu\nu}a.\label{eq:CSid}
\end{equation}

To finally show the conservation of $T^{\mn}$ in flat Minkowski space, we also need the Ward identity 
\be
\nabla^{(b)}_\mu\mathcal{T}^{\mu\nu}=\frac{\beta}{Q_s^4}\mathcal{H}\nabla^{(b)\nu} h+\frac{\alpha}{Q_s^4}\mathcal{A}\nabla^{(b)\nu} a
\ee
in the following form
\begin{equation}
\partial_\mu\wT^{\mu\nu}=\frac{\beta}{Q_s^4}\wH g^{(b) \mu\nu}\partial_\mu h+\frac{\alpha}{Q_s^4}\wA g^{(b) \mu\nu}\partial_\mu a-\Gamma^\nu_{\mu\sigma}\wT^{\mu\sigma},
\end{equation}
where the Christoffel symbol is
\begin{equation}
\Gamma^\nu_{\mu\sigma}=\frac{\gamma}{Q_s^4}g^{(b) \nu\tau}\left(\partial_{(\mu} t_{\sigma)\tau}-\frac{1}{2}\partial_{\tau} t_{\mu\sigma}\right).
\end{equation}
Putting everything together we finally arrive at 
\begin{eqnarray}
\partial_\mu T^{\mu\nu}&=&\frac{1}{N_c}\mathrm{Tr}[({EOMs})^\beta F^{\nu}_{\ \beta}]-\frac{\beta}{Q_s^4}\partial^\nu h\wH-\frac{\alpha}{Q_s^4}\partial^\nu a\wA\nonumber\\
&\ &+\partial_\mu\wT^{\mu\nu}
-\frac{\gamma}{2Q_s^4}\wT^{\alpha\beta}\partial^\nu t_{\alpha\beta}+\frac{\gamma}{Q_s^4}\partial_\mu(\wT^{\mu\sigma}t_\sigma^{\ \nu})\nonumber\\
&=&\frac{1}{N_c}\mathrm{Tr}[({EOMs})^\beta F^{\nu}_{\ \beta}]+\left(g^{(b) \mu\nu}-\eta^{\mu\nu}+t^\nu_{\alpha} g^{(b) \alpha\mu}\right)\left(\frac{\beta}{Q_s^4}\partial_\mu h\wH+\frac{\alpha}{Q_s^4}\partial_\mu a\wA\right)\nonumber\\
&\ &-\left[\Gamma^\nu_{\mu\sigma}+\frac{\gamma}{Q_s^4}t^\nu_{\alpha}\Gamma^\alpha_{\mu\sigma}-\frac{\gamma}{Q_s^4}\eta^{\nu\tau}\left(\partial_{(\mu} t_{\sigma)\tau}-\frac{1}{2}\partial_{\tau} t_{\mu\sigma}\right)\right]\wT^{\mu\sigma},\label{eq:conservation}
\end{eqnarray}
where we used the Ward identity in the second equality
and where $(\text{\it EOMs})^\beta=0$ represents the equations of motion (\ref{eom}).
Using matrix notation the inverse metric can be written as ($\eta^{(-1)}=\eta$)
\begin{equation}
g_{(b)}^{(-1)}=\left[\eta(\mathds{1}+\frac{\gamma}{Q_s^4} t )\right]^{(-1)}=\left[(\mathds{1}+\frac{\gamma}{Q_s^4} t )\right]^{(-1)}\eta,
\end{equation}
and therefore
\begin{equation}
\eta^{\nu\tau}=g^{(b) \nu\tau}+\frac{\gamma}{Q_s^4}t^\nu_\alpha g^{(b) \alpha\tau}.\label{eq:metricid}
\end{equation}
Inserting Eq. \eqref{eq:metricid} in \eqref{eq:conservation} we obtain that on-shell
\begin{equation}
\partial_\mu T^{\mu\nu}=0.
\end{equation}

\section{Computation of the IR-CFT energy-momentum tensor}\label{hol}

In this appendix we give some details of the computation of the result (\ref{eq:emt})
for the IR-CFT energy-momentum tensor $\hat{\mathcal T}^{\mn}$.
We will follow the standard procedure to obtain 
it from the asymptotic expansion in the holographic (radial) coordinate of the metric in Fefferman-Graham gauge \cite{deHaro:2000xn}. The transformation to the Fefferman-Graham gauge will affect the coordinates $r$ and $v$, $r\rightarrow r(\rho,t)$ and $v\rightarrow v(\rho,t)$, where $\rho$ denotes the holographic radial coordinate with the boundary located at $\rho=0$  and $t$ being the time coordinate of the boundary theory. The asymptotic expansion of the transformation reads
\begin{eqnarray}\label{expansion}
r(\rho,t)&=&\sum_{i=0}r_{\ms{(i)}}(t)\rho^{i-1},\nonumber\\
v(\rho,t)&=&\sum_{i=0}v_{\ms{(i)}}(t)\rho^{i},
\end{eqnarray}
which we plug into \eqref{eq:AdSSS} and solve for the coefficients by  demanding  that $g_{\rho\rho}=1/\rho^2$ and $g_{\rho t}=0$ to all orders in $\rho$. In addition we have to impose the condition that the induced metric on a 
constant-$\rho$ slice is regular at leading order. At leading order these conditions lead to $v_{\ms{(1)}}=0$ and to the metric
\begin{equation}\label{eq:AdSSSFG}
\mathrm{d}s^2=\frac{1}{\rho^2}\left[\mathrm{d}\rho^2+r_{\ms{(0)}}(t)^2\left(-v_{\ms{(0)}}'(t)^2\mathrm{d}t^2+\mathrm{d}x^2+\mathrm{d}y^2+\mathrm{d}z^2\right)\right]+\mathcal{O}(\rho^0). 
\end{equation}
To obtain this metric we have only   employed  a  coordinate transformation and therefore it  also solves  Einsteins equation with  negative cosmological constant.  Put differently, the transformation simply changes the rods  and clocks of an asymptotic observer in a time dependent way.
Matching this form of the metric  to the boundary condition \eqref{eq:bc1} gives 
\begin{eqnarray}\label{coeff}
r_{\ms{(0)}}(t)=\sqrt{1+(\gamma/Q_s^4)p(t)}\nonumber\\
v_{\ms{(0)}}'(t)=\sqrt{\frac{1-(\gamma/Q_s^4)3p(t)}{1+(\gamma/Q_s^4)p(t)}}~.
\end{eqnarray}
This is exactly what we wanted, because now we have a bulk geometry that is solely determined by the pressure of the Yang-Mills theory.
To construct the full holographic energy-momentum tensor one also has to solve for the higher-order coefficients in the expansion (\ref{expansion}), which can be expressed in terms of $r_{(0)},~v_{(0)}$ and derivatives thereof.

The energy-momentum tensor density is obtained from the well known formula \cite{deHaro:2000xn}
\begin{equation}
\mathcal{T}^{\mu\nu}=\frac{N_c^2}{2\pi^2}\left\{g^{\ms{(4)}\mu\nu}-\frac{1}{2}g^{\ms{(2)}\mu\sigma}g^{\ms{(2)}\nu}_{\ \sigma}+\frac{1}{4}g^{\ms{(2)}\mu\nu}\mathrm{tr}g^{\ms{(2)}}-\frac{1}{8}g^{\ms{(b)}\mu\nu}\left[\left(\mathrm{tr}g^{\ms{(2)}}\right)^2-g^{\ms{(2)}\sigma\tau}g^{\ms{(2)}}_{\sigma\tau}\right]\right\},
\end{equation}
where $g^{(2)}$ is the $\mathcal{O}(\rho^0)$-term and $g^{(4)}$ the  $\mathcal{O}(\rho^2)$-term in the Fefferman-Graham expansion (\ref{asymp}) of the bulk metric.

\begin{figure}[t]
\begin{center}
\includegraphics[scale=0.7]{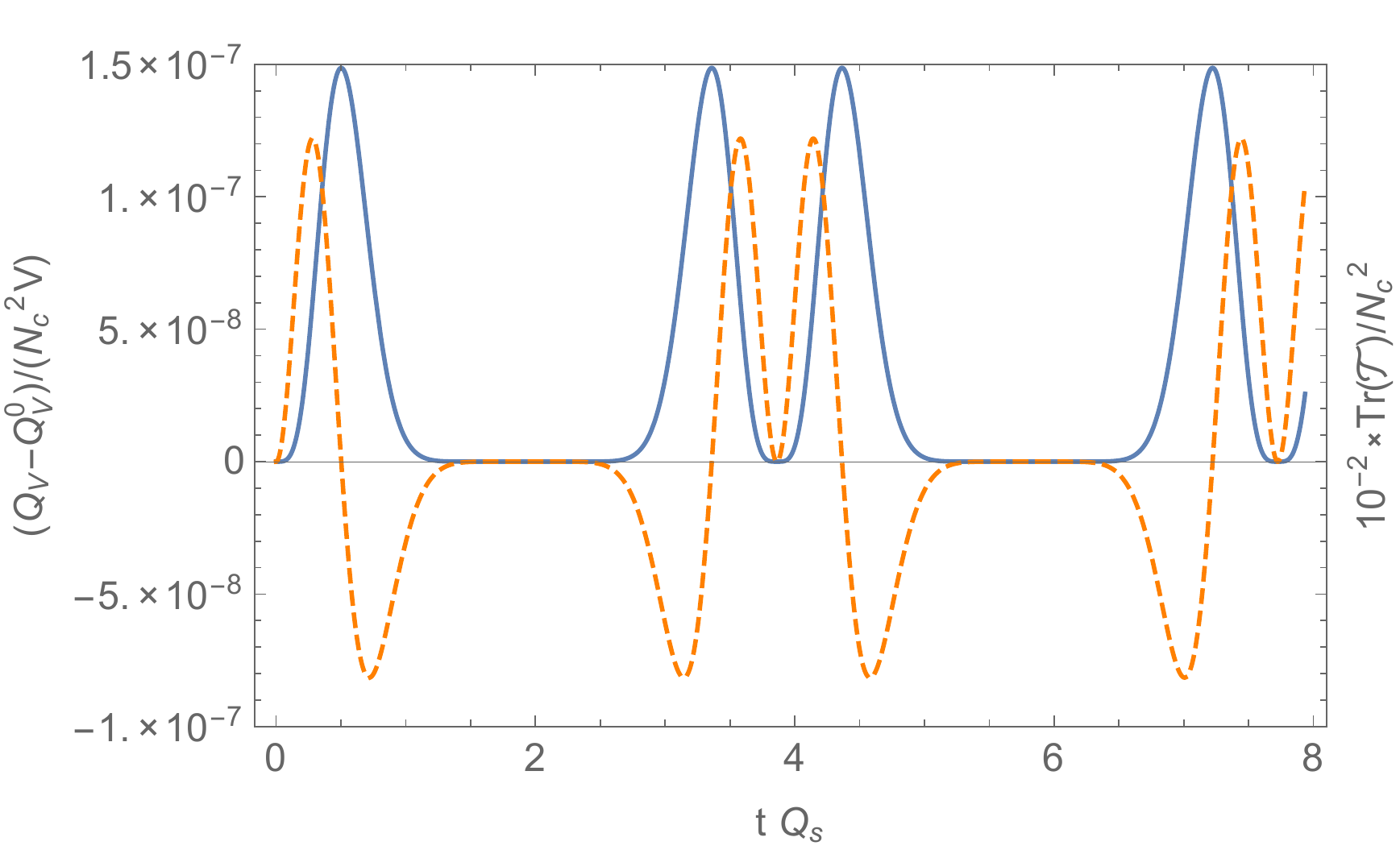}
\caption{The solid (blue) line shows the time dependent (anomalous) contributions in \eqref{eq:canonicalcharge2} ($Q_V^0$ denoting the charge of the AdS-Schwarzschild black hole with flat boundary) for the case (b) of Fig.~\ref{fig:numericalsolution}. The dashed (orange) line shows the evolution of $\mathrm{Tr}\,\mathcal{T}$.\label{fig:canonicalcharge}}
\end{center}
\end{figure}

From the purely gravitational perspective it is interesting to calculate the canonical charges associated with the boundary condition preserving transformations \cite{Henneaux:1985tv,Castellani:1981us}. Since the boundary metric is conformally flat, the asymptotic symmetries are  still given by the conformal group $SO(2,4)$.
If the boundary were flat, the vector field generating infinitesimal time translations would be $\xi_{\ms{(t)}}^\mu=(1,0,0,0)^T$. This in turn can be associated with a canonical charge that is interpreted as the total gravitational energy contained in the bulk of the space-time. In general, given a generator of a boundary condition preserving transformation $\xi_{\ms{(i)}}^\mu\ \in\ SO(2,4)$ the associated canonical charge reads
\begin{equation}
Q_V[\xi_{\ms{(i)}}]=\int_V\mathrm{d}^3x\overline{\mathcal{T}}^{t}_\mu\xi_{\ms{(i)}}^\mu,\label{eq:canonicalcharge1}
\end{equation}
and is preserved in time up to effects due to the conformal anomaly. In \eqref{eq:canonicalcharge1} we introduced a regulating spatial coordinate volume $V=\Delta x\Delta y\Delta z$, since the spatial hypersurfaces in the boundary space-time are non-compact. In the case at hand the generator of time translations is promoted to a conformal Killing vector field of the transformed metric $\xi_{\ms{(t)}}^\mu=(1/v_{\ms{(0)}}'(t),0,0,0)^T$. The charge associated with this becomes
\begin{equation}\label{eq:canonicalcharge2}
Q_V[\xi_{\ms{t}}]=V\frac{3N_c^2}{16\pi^2}\left(4 c+\frac{r_{\ms{(0)}}'(t)^4}{r_{\ms{(0)}}(t)^4 v_{\ms{(0)}}'(t)^4}\right),
\end{equation}
where the first contribution is identical to the canonical charge with flat boundary conditions. The second contribution is always positive and vanishes for $p'(t)=0$. This behavior is displayed in Fig.\ \ref{fig:canonicalcharge}.

\end{appendix}

\bibliographystyle{JHEP} 

\bibliography{semiholo}

\end{document}